\documentclass[aps,prd,10pt,notitlepage,nofootinbib,superscriptaddress]{revtex4-1}

\usepackage[utf8]{inputenc}
\usepackage{amsmath}
\usepackage{amssymb}
\usepackage{amsfonts}
\usepackage{newtxtext,newtxmath}
\usepackage{bm}
\usepackage{graphicx}
\usepackage[usenames,dvipsnames]{xcolor}
\usepackage{color}
\usepackage[colorlinks=true,linkcolor=blue,urlcolor=blue,citecolor=blue]{hyperref}
\usepackage{slashed}
\usepackage[english]{babel}
\usepackage{dcolumn}
\usepackage{blindtext}
\usepackage{epsfig}
\usepackage{pifont}
\usepackage{dsfont,mathrsfs}
\usepackage{cancel}
\usepackage{bigints}
\usepackage{accents}
\usepackage{soul}
\usepackage{multirow}
\usepackage{simpler-wick}

\newcommand{\nn}{\nonumber}

\newcommand{\ensembleaverage}[1]{\left\langle#1\right\rangle}

\newcommand{\Ensembleaverage}[1]{\langle#1\rangle}

\newcommand{\FB}[1]{\left(#1\right)}
\newcommand{\fb}[1]{(#1)}
\newcommand{\SB}[1]{\left\{#1\right\}}
\newcommand{\TB}[1]{\left[#1\right]}

\newcommand{\mcTc}{\mathcal{T}_C}
\newcommand{\mcN}{\mathcal{N}}

\newcommand{\scrD}{\mathscr{D}}

\newcommand{\munu}{{\mu\nu}}

\newcommand{\IM}{\text{Im}}

\newcommand{\Tr}{\text{Tr}}

\newcommand{\kpll}{k_\parallel}

\newcommand{\ppll}{p_\parallel}

\newcommand{\kper}{k_\perp}

\newcommand{\gpll}{g_\parallel}
\newcommand{\gper}{g_\perp}

\newcommand{\del}{\partial}

\newcommand{\psibar}{\overline{\psi}}

\begin{document}

\title{Quantum field theoretical structure of electrical conductivity of cold and dense fermionic matter in the presence of a magnetic field}

\author{Sarthak Satapathy}
\email{sarthaks@iitbhilai.ac.in}
\affiliation{Indian Institute of Technology Bhilai, GEC Campus, Sejbahar, Raipur - 492015, Chhattisgarh, India}
\affiliation{School of Physical Sciences, National Institute of Science Education and Research, Bhubaneswar, HBNI, Jatni - 752050, India}
\author{Snigdha Ghosh}
\email{snigdha.physics@gmail.com}
\thanks{Corresponding Author}
\affiliation{Government General Degree College Kharagpur-II, Paschim Medinipur - 721149, West Bengal, India}
\author{Sabyasachi Ghosh}
\email{sabyaphy@gmail.com}
\affiliation{Indian Institute of Technology Bhilai, GEC Campus, Sejbahar, Raipur - 492015, Chhattisgarh, India}

\begin{abstract}
We have gone through a detailed calculation of the two-point correlation function of vector currents at finite density and magnetic field by employing the real time formalism of finite temperature field theory and Schwinger's proper time formalism. With respect to the direction of external magnetic field, the parallel and perpendicular components of electric conductivity for the degenerate relativistic fermionic matter are obtained from the zero momentum limit of the current-current correlator, owing to Kubo formula. Our quantum field theoretical expressions and numerical estimations are compared with the same, obtained from the relaxation time approximation methods of kinetic theory and its Landau quantized extension, which may be called as classical and quantum results respectively. All the results are merged in the classical domain i.e. high density and low magnetic field region but in the remaining (quantum) domain, quantum results carry a quantized information like Shubnikov-de Haas oscillation along density and magnetic field axes. We have obtained completely new quantum field theoretical expression for perpendicular conductivity of degenerate relativistic fermionic matter. Interestingly, our quantum field theoretical calculation provide a new mathematical form of cyclotron frequency with respect to its classical definition, which might require more future research to interpret the phenomena.  
\end{abstract}
\maketitle

%+++++++++++++++++++++++++++++++++++++++++++++++++++++++++++++++++++++++++++++++++++++++++++++++++++++++++++++++++++
\section{Introduction}
An extremely high density and strong magnetic field~\cite{Harding:2006qn} is naturally found in compact stars like white dwarf (WD) and neutron stars (NS), which have 
long been studied as a focus research problem of nuclear and astrophysics sector. These happen to be the dead remnants of massive stars, the cores of which have been 
collapsed during supernovae collision and a complicated layer structure has been formed leading to a compact structure as the leftover. They are not massive enough to 
form black holes because of an incomplete collapse. Based on various studies~\cite{Seiradakis:2004ew,McGill}, we know a gross range from $10^{12}$G~\cite{Seiradakis:2004ew} 
to $10^{15}$G~\cite{McGill} for the surface magnetic fields in NS. With increasing depth the density of matter increases reaching upto $\rho = 2.8 \times 10^{14}gm/cm^3$ 
\cite{Shapiro:1983du}. At this density nucleons cease to exist and the matter is made up of quarks. From Maxwell's equations we know that magnetic flux is conserved. 
This leads to the conclusion that the strength of magnetic field is $> 10^{16}G$ in the interior of neutron stars and magnetars. The strength of magnetic field varies 
depending on the nature of the core. For a core made up of neutrons the magnetic field produced is of the order $10^{18}$ G and for a quark core \cite{Annala:2019puf} 
it is of the order $10^{20}$G \cite{Kharzeev:2012ph}. This magnetic field strength of NS can have Ohmic decay profile, which depends on the electrical conductivity of 
the NS~\cite{Goldreich}. To solve relativistic magnetohydrodynamics equations for simulating magnetized neutron stars or binary star mergers, the electrical conductivity 
of the crustal matter becomes an important input~\cite{Dionysopoulou:2012zv,Palenzuela:2013hu,Palenzuela:2013kra,Dionysopoulou:2015tda,Harutyunyan:2018mpe}. Due to the 
recently observed gravitational wave signal GW170817~\cite{LIGOScientific:2017vwq}, the binary neutron star merger simulation has gained  attention and thus unfolding a 
new  field of research - multimessenger astronomy. In these connections, the microscopic calculation of electrical conductivity in presence of magnetic field might be an 
important research topic. One can find a long history with a long list of references (e.g. few selective Refs.~\cite{mestel_1950,Abrikosov:1964,Canuto:1970,Flowers:1976}) 
for microscopic calculation of electrical conductivity for compact star but those references have not considered the impact of magnetic field. For those calculations at 
finite magnetic field picture, reader can go through the Refs.~\cite{Canuto:1969ybo,Canuto:1970ai,Potekhin:1999ur,Ostgaard:1992pyt,Kerbikov:2014ofa,Baiko:2016nzl,Harutyunyan:2016rxm}, 
among which Refs.~\cite{Potekhin:1999ur,Kerbikov:2014ofa} have only explored the quantum effect or Landau quantization aspect in electrical conductivity expressions.

Similar kind of microscopic calculations~\cite{Nam:2012sg,Hattori:2016lqx,Kurian:2018qwb,Kurian:2017yxj,Feng:2017tsh,Fukushima:2017lvb,Das:2019ppb,Das:2019pqd,Das:2019wjg,Dey:2019axu,Dey:2019vkn,Dash:2020vxk,Ghosh:2019ubc,Samanta:2020wyh} high temperature and low density QCD matter, which can be produced in heavy ion collision (HIC) experiments like relativistic heavy ion collision (RHIC) and large hadron collider (LHC). By increasing the temperature, one can expect hadron to quark phase transition at nearly zero (net) quark/baryon density. This early universe environment can be expected in RHIC or LHC experiments, where a huge magnetic field can also be created in the peripheral collisions. One can expect a super-hot massless quark gluon plasma (QGP) under a strong magnetic field, whose conductivity expression for classical, quantum field theory cases are respectively discussed in Refs.~\cite{Dey:2019axu,Satapathy:2021cjp}. Here, classical terminology is used for the case, where no Landau quantizations have not been considered and calculations are based on relaxation time approximation (RTA) based kinetic theory. One can impose Landau quantizations into RTA expressions to get their quantum expressions~\cite{Ghosh:2019ubc}, but they are not same exactly with Kubo expressions~\cite{Satapathy:2021cjp}, which can be considered as quantum field theoretical expressions. Similar to the super-hot massless QGP at strong magnetic field in RHIC or LHC experiments, the core of the NS~\cite{Annala:2019puf,Baym:2017whm} can have a super-dense masless quark matter with a strong magnetic field, whose classical to quantum estimation has been done in Ref.~\cite{Dey:2021fbo}. In present work, we will explore its quantum field theoretical structure. Previously~\cite{Satapathy:2021cjp}, we have explored the same field theoretical structure in the high temperature and magnetic field domain. Interestingly, the field theoretical structure in the high temperature and magnetic field domain, which is addressed in present work, is showing an oscillatory pattern along magnetic field or density axis. This kind of oscillatory pattern in condense matter field is well known fact in low temperature and strong magnetic field domain and it is popularly called Shubnikov-de Haas (SdH) effect~\cite{Schubnikow:1930.1,Schubnikow:1930.2,Schubnikow:1930.3,Schubnikow:1930.4} or SdH oscillations where it was found that the resistivity oscillates as a function of magnetic field. The SdH effect is a purely quantum mechanical effect. Similar kind of quantized effect probably can be noticed in quark core of NS, facing strong magnetic field. This possibility is indicated by Ref.~\cite{Dey:2021fbo} via RTA methods with Landau quantization, while present article is revealing the same possibility via quantum field theoretical methods, carrying some enriched structure. With the help of Schwinger's proper time methods~\cite{Schwinger:1951nm,Chyi:1999fc,Kuznetsov:2004tb,Kuznetsov:2013sea,Schwartz:2014sze,Calzetta:2008iqa,Toms:2012bra,Inagaki:2003ac,Kuznetsov:2015uca,Ayala:2004dx} and the real time formalism (RTF) of thermal field theory~\cite{Bellac:2011kqa,Mallik:2016anp,Kapusta:2006pm,Mallik:2009pj,Weldon:2007zz,Kobes:1984vb,Lundberg:2020mwu}, we have calculated the vector current-current correlation function involving fermionic fields, whose zero momentum limit, based on Kubo-Zubarev formalism~\cite{zubarev1974nonequilibrium,zubarev1996statistical,Huang:2011dc,Hosoya:1983id}, provide a rich field theoretical expression of conductivity.   

The article is organized as follows. In Sec.~\ref{sec2}, we have addressed the detail calculation of two point function at finite density and magnetic field, which at the end reach to the final expressions of parallel and perpendicular conductivity components of degenerate quark matter. Next in Sec.~\ref{sec3}, we have generated our quantum field theoretical resuluts, which are compared with the classical and quantum results, defined in earlier work. At the end, Sec.~\ref{sec4} summarizes our studies and zoom in the field theoretical ingredients, which is addressed first time here. Some anatomy of calculations are kept in Appendix.

\textbf{Notations:} We use natural unit in which $\hbar = c = k_B = 1$. The metric tensor has signature $g^\munu = \text{diag}(1,-1,-1,-1)$. With respect to the $\hat{z}$-direction, we decompose any four vector $k^\mu$ as $k=(\kpll+\kper)$ where $\kpll^\mu=\gpll^\munu k_\nu$ and $\kper^\mu=\gper^\munu k_\nu$ in which the corresponding decomposition of the metric tensor is $g^\munu=(\gpll^\munu+\gper^\munu)$ with  $\gpll^\munu=\text{diag}(1,0,0,-1)$ and $\gper^\munu=\text{diag}(0,-1,-1,0)$. Note that, $\kper^2 = -(k_x^2+k_y^2)<0$ is a space-like vector.

%++++++++++++++++++++++++++++++++++++++++++++++++++++++++++++++++++++++++++++++++++++++++++++++++++++++++++++++++++++++++
\section{Formalism} \label{sec2}
Let us consider a system of non-interacting charged fermions of mass $m$ and charge $e>0$ in a constant background magnetic field $\bm{B}=B\hat{z}$. The system is described by the Lagrangian (density): 
\begin{eqnarray}
\mathcal{L}_\text{Dirac} = \psibar \TB{  i \gamma^\mu D_\mu - m } \psi
\label{eq.lagrangian}
\end{eqnarray} 
where $\psi, \overline{\psi}$ are the conjugate Dirac fields, $D_{\mu} = \partial_{\mu} + ie\widetilde{A}_{\mu}$ is the gauge covariant derivative owing to the minimal electromagnetic coupling, and, $\widetilde{A}^{\mu} = A^{\mu} + A^{\mu}_\text{ext}$. Here, $A^{\mu}$ is the dynamical photon field and $A^{\mu}_\text{ext}$ is the classical four-potential arising from the background magnetic field. The calculation will be valid even if the fermion field $\psi$ is a multiplet.
\begin{figure}[h]
\begin{center}
\includegraphics[angle=0,scale=0.4]{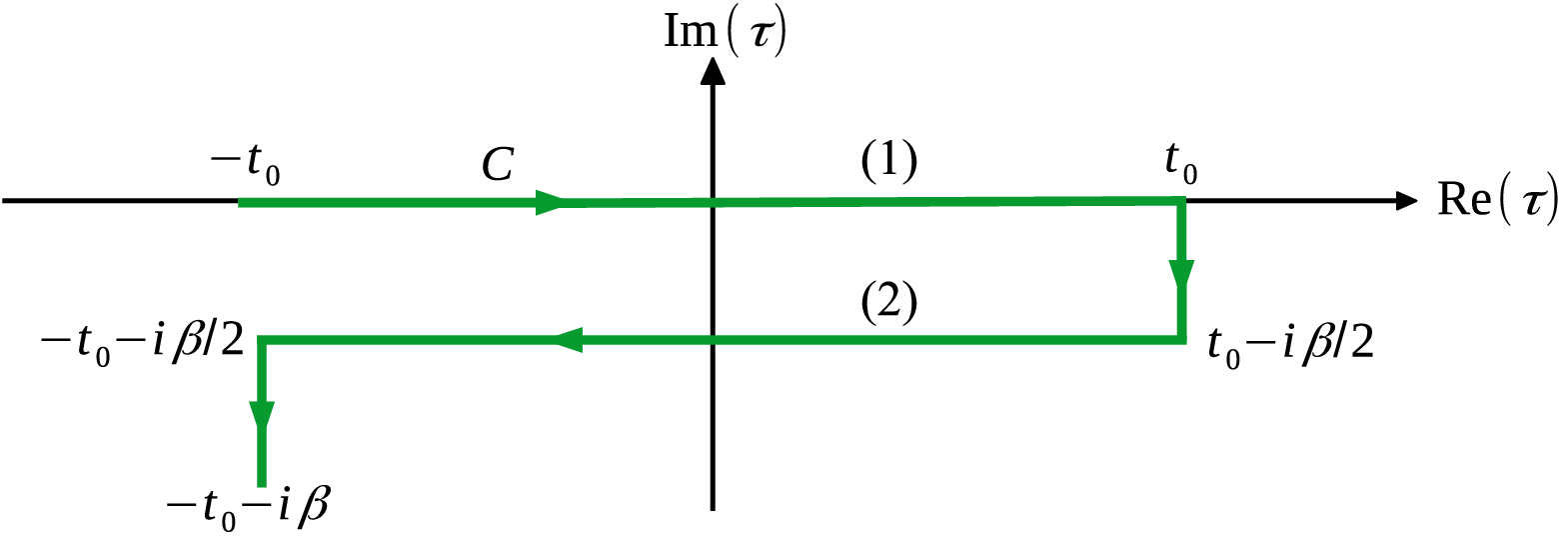}
\end{center}
\caption{The symmetric Schwinger-Keldysh contour $C$ in the complex time ($\tau$) plane used in the RTF with $t_0\to\infty$ and inverse temperature $\beta = 1/T$. The two horizontal segments of the contour are labeled as  `(1)' and `(2)' respectively.}
\label{fig.contour}
\end{figure}

To calculate the electrical conductivity ($\sigma$), we first require the in-medium thermo-dense-magnetic spectral function $\rho^{\mu\nu}(q)$, calculated from the ensemble average of the two-point correlation function of local vector currents, given by~\cite{Satapathy:2021cjp}
\begin{eqnarray}
\rho^{\mu\nu}(q) = \tanh\FB{\frac{q^0}{2T}} \IM~i\int d^4x e^{iq \cdot x} \Ensembleaverage{\mcTc J^{\mu}(x)J^{\nu}(0)}_{11}^B ~,
\label{eq.sf.1}
\end{eqnarray}
where $J^{\mu}(x)$ denotes the conserved Noether's current corresponding to the $U(1)$ global gauge symmetry of the Lagrangian in Eq.~\eqref{eq.lagrangian}, $\mcTc$ is the time-ordering with respect to the symmetric Schwinger-Keldysh contour $C$ in the complex time plane (shown in Fig.~\ref{fig.contour}) as used in the RTF of finite temperature field theory~\cite{Bellac:2011kqa,Mallik:2016anp,Kapusta:2006pm,Weldon:2007zz,Mallik:2009pj,Kobes:1984vb,Inagaki:2003ac}, and, the subscript $`11'$ refers to the fact that the two points are on the real horizontal segment of $C$. The explicit form of the current $J^{\mu}(x)$ reads
\begin{eqnarray}
J^{\mu}(x) = e \psibar \gamma^\mu\psi.
\label{eq.current}
\end{eqnarray}

Using Eq.~\eqref{eq.current}, we now calculate the two-point correlation function $\Ensembleaverage{\mcTc J^{\mu}(x)J^{\nu}(0)}_{11}^B$, the details of which are provided in Appendix~\ref{appendix.A}. From Eq.~\eqref{eq.A1.corr.2}, we read off the final result as 
\begin{eqnarray}
\ensembleaverage{\mcTc J^\mu(x)J^\nu(0)}^B_{11} &=& 
-\sum_{l=0}^{\infty} \sum_{n=0}^{\infty}\int\!\!\!\int\!\!\!\frac{d^4p}{(2\pi)^4}\frac{d^4k}{(2\pi)^4} e^{-ix\cdot(p-k)}
D_{11}(\ppll,m_n)D_{11}(\kpll,m_l) 
\mcN_{ln}^\munu(k,p)
\label{eq.corr.1}
\end{eqnarray}
where $m_l = \sqrt{ m^2 + 2leB }$ is the {\it effective Landau level dependent mass}, and, $D_{11}$ and $\mathcal{N}_{ln}^{\mu\nu}$ are defined respectively in Eqs.~\eqref{eq.A1.D11} and \eqref{eq.A1.N.2}. On substituting Eq.~\eqref{eq.corr.1} into Eq.~\eqref{eq.sf.1} and simplifying, we obtain after a bit algebra 
\begin{eqnarray}
\rho^{\mu\nu}(q) = -\tanh\FB{\frac{q^0}{2T}}\IM~i \int\!\!\!\frac{d^4k}{(2\pi)^4}\sum_{n=0}^{\infty}\sum_{l=0}^{\infty}
D_{11}(k_\parallel,m_l)D_{11}(p_\parallel = q_\parallel +k_\parallel,m_n)\mathcal{N}^{\mu\nu}_{ln}(k,p=q+k).
\label{eq.sf.2}
\end{eqnarray}
Again substituting the expression of $D_{11}$ from Eq.~\eqref{eq.A1.D11} into Eq.~\eqref{eq.sf.2} and performing the $dk_0d^2\kper$ integral, we get
\begin{eqnarray}
\rho^{\mu\nu}(q) = \tanh\FB{\frac{q^0}{2T}}\pi && \sum_{l=0}^{\infty}\sum_{n=0}^{\infty} \int_{-\infty}^{\infty}\frac{dk_z}{2\pi}\frac{1}{4\omega_{kl}\omega_{pn}} \nn \\
&& \times \Big[
\big\{1 - f_-(\omega_{kl}) - f_+(\omega_{pn}) + 2f_-(\omega_{kl})f_+(\omega_{pn})\big\} 
\mathcal{\widetilde{N}}^{\mu\nu}_{ln}(k_0=-\omega_{kl})\delta(q_0-\omega_{kl}-\omega_{pn}) \nn \\
&& +~ \big\{1 - f_+(\omega_{kl}) - f_-(\omega_{pn}) + 2f_+(\omega_{kl})f_-(\omega_{pn})\big\} 
\mathcal{\widetilde{N}}^{\mu\nu}_{ln}(k_0=\omega_{kl})\delta(q_0+\omega_{kl}+\omega_{pn}) \nn \\
&& +~ \big\{ - f_-(\omega_{kl}) - f_-(\omega_{pn}) + 2f_-(\omega_{kl})f_-(\omega_{pn})\big\} 
\mathcal{\widetilde{N}}^{\mu\nu}_{ln}(k_0=-\omega_{kl})\delta(q_0-\omega_{kl}+\omega_{pn}) \nn \\
&& +~ \big\{ - f_+(\omega_{kl}) - f_+(\omega_{pn}) + 2f_+(\omega_{kl})f_+(\omega_{pn})\big\} 
\mathcal{\widetilde{N}}^{\mu\nu}_{ln}(k_0=\omega_{kl})\delta(q_0+\omega_{kl}-\omega_{pn}) 
\Big]
\label{eq.sf.3}
\end{eqnarray}
where $\omega_{kl}= \sqrt{k_z^2+m_l^2}$, $\omega_{pn} = \sqrt{(p_z+q_z)^2 + m_n^2}$,  $f_\pm(\omega)=\TB{e^{(\omega\mp \mu )/T}+1}^{-1}$ are the Fermi-Dirac thermal distribution functions, and, $\mathcal{\widetilde{N}}_{ln}^{\mu\nu}(\kpll) = {\displaystyle \int}\!\!\dfrac{d^2\kper}{(2\pi)^2}\mathcal{N}_{ln}^{\mu\nu}(k,k)$ can be read off from Eq.~\eqref{eq.A1.Ntil.2} as
\begin{eqnarray}
\mathcal{\widetilde{N}}_{ln}^{\mu\nu}(\kpll) = e^2  \frac{eB}{\pi} \Big[
-4eBn\delta_{l-1}^{n-1} \gpll^\munu
- \FB{\delta_{l}^{n}+\delta_{l-1}^{n-1}}\SB{ 2k_\parallel^{\mu}k_\parallel^{\nu}-\gpll^{\mu\nu}(k_\parallel^2-m^2)} 
+ \FB{\delta_{l}^{n-1} + \delta_{l-1}^{n}}\FB{k_\parallel^2-m^2}g_{\perp}^{\mu\nu}
\Big].
\label{eq.Ntil.1}
\end{eqnarray}

The RHS of Eq.~\eqref{eq.sf.3} contains four Dirac delta function and they give rise to the branch cuts of the spectral function in the complex energy ($q^0$) plane. First two delta functions are called the Unitary cuts whereas the last two delta functions are termed as the Landau cuts. In order to calculate transport coefficients, we need to take the static (long wavelength) limit i.e $\bm{q}=\bm{0} , q^0\to 0$ of the spectral function in Eq.~\eqref{eq.sf.3}, so that only the Landau cuts contribute and we are left with
\begin{eqnarray}
\rho^{\mu\nu}( q_0, \bm{q}=\bm{0}) &=& \tanh\FB{\frac{q^0}{2T}}\pi\sum_{l=0}^{\infty}\sum_{n=0}^{\infty}\int_{-\infty}^{\infty}\frac{dk_z}{2\pi}\frac{1}{4\omega_{kl}\omega_{kn}}
\nn \\ 
&& \times \Big[
\big\{ - f_-(\omega_{kl}) - f_-(\omega_{kn}) + 2f_-(\omega_{kl})f_-(\omega_{kn})\big\} 
\mathcal{\widetilde{N}}^{\mu\nu}_{ln}(k_0=-\omega_{kl})\delta(q_0-\omega_{kl}+\omega_{kn}) \nn \\
&& +~ \big\{ - f_+(\omega_{kl}) - f_+(\omega_{kn}) + 2f_+(\omega_{kl})f_+(\omega_{kn})\big\} 
\mathcal{\widetilde{N}}^{\mu\nu}_{ln}(k_0=\omega_{kl})\delta(q_0+\omega_{kl}-\omega_{kn}) 
\Big] \nn \\
&=& \lim\limits_{\Gamma\to0} \tanh\FB{\frac{q^0}{2T}}\sum_{l=0}^{\infty}\sum_{n=0}^{\infty}\int_{-\infty}^{\infty}\frac{dk_z}{2\pi}\frac{1}{4\omega_{kl}\omega_{kn}}
\nn \\
&& \times \Big[
\big\{ - f_-(\omega_{kl}) - f_-(\omega_{kn}) + 2f_-(\omega_{kl})f_-(\omega_{kn})\big\} 
\mathcal{\widetilde{N}}^{\mu\nu}_{ln}(k_0=-\omega_{kl}) \frac{\Gamma}{\Gamma^2+(q_0 - \omega_{kl} + \omega_{kn})^2}  \nn \\
&& +~ \big\{ - f_+(\omega_{kl}) - f_+(\omega_{kn}) + 2f_+(\omega_{kl})f_+(\omega_{kn})\big\} 
\mathcal{\widetilde{N}}^{\mu\nu}_{ln}(k_0=\omega_{kl}) \frac{\Gamma}{\Gamma^2+(q_0 + \omega_{kl} - \omega_{kn})^2}  
\Big]
\end{eqnarray}
where, in the last step, we have used the Breit-Wigner representation of the Dirac-delta function $\delta(x) = \frac{1}{\pi} \lim\limits_{\Gamma\to0}\IM\FB{\frac{1}{x-i\Gamma}} =  \lim\limits_{\Gamma\to0} \FB{ \frac{\Gamma}{\Gamma^2+x^2}}$. 
The conductivity tensor $\sigma^\munu$ in the presence of external magnetic field is obtained in the Kubo formalism by taking the zero-momentum limit of $\rho^\munu/q^0$ or alternatively by using the L'Hospital's rule as 
\begin{eqnarray}
\sigma^{\mu\nu}(T,\mu,B) = \frac{\partial\rho^{\mu\nu}}{\partial q_0}\Big|_{\bm{q}= \bm{0},q_0 \to 0} 
&=& \lim\limits_{\Gamma\to 0} \sum_{l=0}^{\infty}\sum_{n=0}^{\infty}\frac{1}{2T}\int_{-\infty}^{\infty}\frac{dk_z}{2\pi}\frac{1}{4\omega_{kl}\omega_{kn}}
\frac{\Gamma}{\Gamma^2 + (\omega_{kl} -\omega_{kn})^2} 
\nn \\ 
&& \times \Big[
\big\{ - f_-(\omega_{kl}) - f_-(\omega_{kn}) + 2f_-(\omega_{kl})f_-(\omega_{kn})\big\} 
\mathcal{\widetilde{N}}^{\mu\nu}_{ln}(k_0=-\omega_{kl})   \nn \\
&& +~ \big\{ - f_+(\omega_{kl}) - f_+(\omega_{kn}) + 2f_+(\omega_{kl})f_+(\omega_{kn})\big\} 
\mathcal{\widetilde{N}}^{\mu\nu}_{ln}(k_0=\omega_{kl})  
\Big]
\label{eq.cond.tensor.1}
\end{eqnarray}

Since our focal zone is the core of NS, where an extreme relativistic degenerate (super-dense massless) fermionic matter can be expected, so we will impose the $T\to 0$ limit to Eq.~\eqref{eq.cond.tensor.1}. In order to take $T\to0$ limit to Eq.~\eqref{eq.cond.tensor.1}, we first rewrite it as
\begin{eqnarray}
\sigma^{\mu\nu}(T,\mu,B) &=& -\lim\limits_{\Gamma\to 0} \sum_{l=0}^{\infty}\sum_{n=0}^{\infty}\frac{1}{2}\int_{-\infty}^{\infty}\frac{dk_z}{2\pi}\frac{1}{4\omega_{kl}\omega_{kn}}
\frac{\Gamma}{\Gamma^2 + (\omega_{kl} -\omega_{kn})^2} 
\nn \\ 
&& \times \Big[
\Big\{ \frac{f_-(\omega_{kl})}{f_-(\omega_{kn})}\frac{\del}{\del\mu}f_-(\omega_{kn}) + \frac{f_-(\omega_{kn})}{f_-(\omega_{kl})}\frac{\del}{\del\mu}f_-(\omega_{kl}) \Big\} 
\mathcal{\widetilde{N}}^{\mu\nu}_{ln}(k_0=-\omega_{kl})   \nn \\
&& +~ \Big\{ \frac{f_+(\omega_{kl})}{f_+(\omega_{kn})}\frac{\del}{\del\mu}f_+(\omega_{kn}) + \frac{f_+(\omega_{kn})}{f_+(\omega_{kl})}\frac{\del}{\del\mu}f_+(\omega_{kl}) \Big\} 
\mathcal{\widetilde{N}}^{\mu\nu}_{ln}(k_0=\omega_{kl})  
\Big].
\label{eq.cond.tensor.2}
\end{eqnarray}
In the zero temperature limit, the thermal distribution function will behave like the Heaviside step function $\Theta(x)$ as
\begin{eqnarray}
\lim\limits_{T\to 0} f_{\pm}(\omega) = \lim\limits_{T\to 0}\frac{1}{e^{(\omega\mp\mu)/T} + 1} = \Theta(-\omega \pm \mu).
\label{eq.step.1}
\end{eqnarray}
where, the step function is defined as 
\begin{eqnarray}
\Theta (x) = \begin{cases}
1 \text{~~if~~} x>0, \\
1/2 \text{~~if~~} x=0, \\
0 \text{~~if~~} x<0.
\end{cases}
\label{eq.step.2}
\end{eqnarray}
Using Eqs.~\eqref{eq.step.1} and \eqref{eq.step.2} in Eq.~\eqref{eq.cond.tensor.2}, we get after some simplifications
\begin{eqnarray}
\sigma^{\mu\nu}(T\to0,\mu,B) &=& -\lim\limits_{\Gamma\to 0} \sum_{l=0}^{\infty}\sum_{n=0}^{\infty}\int_{-\infty}^{\infty}\frac{dk_z}{2\pi}\frac{1}{4\omega_{kl}\omega_{kn}}
\frac{\Gamma}{\Gamma^2 + (\omega_{kl} -\omega_{kn})^2} 
\nn \\ 
&& \times \Big[
\big\{ \Theta(-\mu-\omega_{kl})\delta(-\mu-\omega_{kn}) + \Theta(-\mu-\omega_{kn})\delta(-\mu-\omega_{kl}) \big\} 
\mathcal{\widetilde{N}}^{\mu\nu}_{ln}(k_0=-\omega_{kl})   \nn \\
&& +~ \big\{ \Theta(\mu-\omega_{kl})\delta(\mu-\omega_{kn}) + \Theta(\mu-\omega_{kn})\delta(\mu-\omega_{kl}) \big\} 
\mathcal{\widetilde{N}}^{\mu\nu}_{ln}(k_0=\omega_{kl})  
\Big].
\label{eq.cond.tensor.3}
\end{eqnarray}
Finally considering the fermion chemical potential $\mu>0$, the antiparticle distributions will not contribute to the spectral function. Even at low $T$ and high density region, one can safely discard the sub-leading anti-particle contribution to the spectral functions. Thus, ignoring the anti-particle part in Eq.~\eqref{eq.cond.tensor.3} we get, 
\begin{eqnarray}
\sigma^{\mu\nu}(T\to0,\mu>0,B) = -\lim\limits_{\Gamma\to 0} \sum_{l=0}^{\infty}\sum_{n=0}^{\infty}\int_{-\infty}^{\infty}\frac{dk_z}{2\pi}\frac{1}{4\omega_{kl}\omega_{kn}}
\frac{\Gamma}{\Gamma^2 + (\omega_{kl} -\omega_{kn})^2} \mathcal{\widetilde{N}}^{\mu\nu}_{ln}(k_0=\omega_{kl}) \nn \\ 
\times \big\{ \Theta(\mu-\omega_{kl})\delta(\mu-\omega_{kn}) + \Theta(\mu-\omega_{kn})\delta(\mu-\omega_{kl}) \big\} .
\label{eq.cond.tensor.4}
\end{eqnarray}

In the presence of background magnetic field, all the transport coefficients including the electrical conductivity $\sigma^{\mu\nu}$ becomes anisotropic. The electrical conductivity parallel ($\parallel$) and perpendicular ($\perp$) to the magnetic field direction can be calculated from 
\begin{eqnarray}
\sigma_\parallel &=& b^{\alpha}b^{\beta}\Delta_{\alpha\mu}\Delta_{\beta\nu} \sigma^{\mu\nu},  \label{eq.sigma.pll.1} \\
\sigma_\perp &=& -\frac{1}{2}\Xi^{\alpha\beta}\Delta_{\alpha\mu}\Delta_{\beta\nu} \sigma^{\mu\nu}, \label{eq.sigma.perp.1}
\end{eqnarray}
where, $\Delta^{\mu\nu} = (g^{\mu\nu}-u^{\mu}u^{\nu})$ is a projector orthogonal to $u^\mu$, $b^\mu = \frac{1}{2B}\varepsilon^{\mu\nu\alpha\beta}F_{\nu\alpha}u_{\beta}$, $F_{\mu\nu} = \FB{\partial_{\mu}A_\nu^\text{ext} - \partial_{\nu}A_\mu^\text{ext}}$ is the electromagnetic field strength tensor, and, $\Xi^{\mu\nu} = (\Delta^{\mu\nu} + b^{\mu}b^{\nu})$. 
Here, $u^\mu$ is the medium four-velocity. In the Local Rest Frame (LRF) of the medium $u^\mu_\text{LRF}\equiv(1,\bm{0})$ and $b^{\mu}_\text{LRF} \equiv (0,\hat{\bm{z}})$.
It can be noticed that, Eqs.~\eqref{eq.sigma.pll.1} and \eqref{eq.sigma.perp.1} also yield the following expressions of the parallel and perpendicular conductivities in terms of different components of the conductivity tensor:
\begin{eqnarray}
\sigma_\parallel &=& \sigma^{33}, \label{eq.sigma.pll.2} \\
\sigma_\perp &=& \frac{1}{2}\FB{\sigma^{11}+\sigma^{22}} \label{eq.sigma.perp.2}
\end{eqnarray}
which are easier to understand in terms of physical interpretation.

Till now, $\Gamma$ in Eq.~\eqref{eq.cond.tensor.4} is an infinitesimal parameter corresponding to the Breit-Wigner representation of the Dirac delta function. Since we have taken a system of non-interacting particles, the transport coefficients should diverge (like the case of an ideal gas) and is apparent from Eq.~\eqref{eq.cond.tensor.4} in the limit $\Gamma\to0$. In order to get non-divergent values of the electrical conductivity, we must consider finite value of $\Gamma>0$ which corresponds to switching on the interactions among the particles, thus allowing dissipation in the medium. $\Gamma$ can be identified as the thermal width (or the inverse of relaxation time $\tau_c$). The thermal width $\Gamma$ can be calculated microscopically from the interaction Lagrangian which involves the estimations of the scattering cross sections (decay rates) among (of) the constituent particles in the presence of magnetic field~\cite{Ghosh:2014qba,Gangopadhyaya:2016jrj,Rahaman:2017sby,Ghosh:2017njg,Ghosh:2018nqi,Ghosh:2019vyw}. In present work, we will keep $\Gamma$ as an input parameter and take it of order of QCD scale $\Gamma\approx 10^2$ MeV or its inverse time scale (the relaxation time) $\tau_c\approx 1$ fm. Here our focal interest will be to see the general structure of magneto-thermodynamical phase-space in electrical conduction for any degenerate and massless Dirac fluid. In the rest of the calculation, we thus continue with finite value of $\Gamma$. Substituting Eq.~\eqref{eq.cond.tensor.4} into Eqs.~\eqref{eq.sigma.pll.1} and \eqref{eq.sigma.perp.1}, we get
\begin{eqnarray}
\sigma_{\parallel,\perp}(\mu,B) = - \sum_{l=0}^{\infty}\sum_{n=0}^{\infty}\int_{-\infty}^{\infty}\frac{dk_z}{2\pi}\frac{1}{4\omega_{kl}\omega_{kn}}
\frac{\Gamma}{\Gamma^2 + (\omega_{kl} -\omega_{kn})^2} \mathcal{\widetilde{N}}^{\parallel,\perp}_{ln}  
\big\{ \Theta(\mu-\omega_{kl})\delta(\mu-\omega_{kn}) + \Theta(\mu-\omega_{kn})\delta(\mu-\omega_{kl}) \big\} 
\label{eq.cond.tensor.5}
\end{eqnarray}
where, 
\begin{eqnarray}
\mathcal{\widetilde{N}}^{\parallel}_{ln} &=& b^{\alpha}b^{\beta}\Delta_{\alpha\mu}\Delta_{\beta\nu} \mathcal{\widetilde{N}}^{\mu\nu}_{ln}(k_0=\omega_{kl}), 
\label{eq.N.pll.1}\\
\mathcal{\widetilde{N}}^{\perp}_{ln} &=& -\frac{1}{2}\Xi^{\alpha\beta}\Delta_{\alpha\mu}\Delta_{\beta\nu} \mathcal{\widetilde{N}}^{\mu\nu}_{ln}(k_0=\omega_{kl}).
\label{eq.N.per.1}
\end{eqnarray}
Substituting Eq.~\eqref{eq.Ntil.1} into Eqs.~\eqref{eq.N.pll.1} and \eqref{eq.N.per.1} and simplifying we get,
\begin{eqnarray}
\mathcal{\widetilde{N}}^{\parallel}_{ln} &=& -2e^2 \frac{eB}{\pi}k_z^2(2-\delta_l^0)\delta_l^n, \label{eq.N.pll.2}\\
\mathcal{\widetilde{N}}^{\perp}_{ln} &=& -2e^2 \frac{(eB)^2}{\pi}l(\delta_l^{n-1}+\delta_{l-1}^n). \label{eq.N.per.2}
\end{eqnarray}
Finally substituting Eqs.~\eqref{eq.N.pll.2} and \eqref{eq.N.per.2} into \eqref{eq.cond.tensor.5} and performing the remaining $dk_z$ integral using the Dirac delta function, we get after long but straightforward algebra the following analytic expressions of the parallel and perpendicular conductivities: 
\begin{eqnarray}
\sigma_\parallel &=& e^2 \FB{\frac{eB}{2\pi^2}}\frac{1}{ \Gamma \mu} \sum_{l=0}^{l_\text{max}}(2-\delta_l^0)\sqrt{\mu^2-m_l^2}\Theta(\mu-m_l), 
\label{eq.sigma.pll.3} \\
\sigma_\perp &=& e^2\FB{\frac{eB}{2\pi^2}} \frac{\Gamma}{\Gamma^2 + (\mu-\sqrt{\mu^2-2eB})^2}\frac{1}{\sqrt{\mu^2-2eB}} 
\sum_{l=1}^{l_\text{max}}\frac{(2l-1)eB}{\sqrt{\mu^2-m_l^2}}\Theta(\mu-m_l), \label{eq.sigma.perp.3}
\end{eqnarray} 
where, 
\begin{eqnarray}
l_\text{max}= \left\lfloor \frac{\mu^2-m^2}{2eB} \right\rfloor
\label{eq.lmax}
\end{eqnarray}
in which the floor function $\lfloor x \rfloor = $ `largest integer less than or equal to $x$'. It can be observed that, the presence of the Kronecker delta functions in Eqs.~\eqref{eq.N.pll.2} and \eqref{eq.N.per.2} have killed one of the double sums of Eq.~\eqref{eq.cond.tensor.5} and the presence of the step functions $\Theta(\mu-m_l)$ in Eqs.~\eqref{eq.sigma.pll.3} and \eqref{eq.sigma.perp.3} have restricted the upper limit of the infinite sum over index $l$ to $l_\text{max}$ given in Eq.~\eqref{eq.lmax}. The step function $\Theta(\mu-m_l)$ also ensures that $\mu$ has to be greater than $2eB$ for a non-vanishing $\sigma_\perp$.

It is also worth noticing that, the presence of the Dirac delta functions in Eq.~\eqref{eq.cond.tensor.4} have made the Fermi momentum along $\hat{\bm{z}}$ direction $k_{zl}^F$ to be quantized as 
\begin{eqnarray}
k_{zl}^F &=& \pm\sqrt{\mu^2-m^2-2leB}~~~;~~l \in \{0,\mathds{Z}^+\} \nn \\
&=& \pm\sqrt{\mu^2-m^2}~,~\pm\sqrt{\mu^2-m^2-2eB}~,\cdots~,~ \pm\sqrt{\mu^2-m^2-2l_\text{max}eB}.
\end{eqnarray}

%~~~~~~~~~~~~~~~~~~~~~~~~~~~~~~~~~~~~~~~~~~~~~~~~~~~~~~~~~~~~~~~~~~~~~~~~~~~~~~~~~~~~~~~~~~~~~~~~~~~~~~~~~~~
%~~~~~~~~~~~~~~~~~~~~~~~~~~~~~~~~~~~~~~~~~~~~~~~~~~~~~~~~~~~~~~~~~~~~~~~~~~~~~~~~~~~~~~~~~~~~~~~~~~~~~~~~~~~
\section{NUMERICAL RESULTS AND DISCUSSIONS} \label{sec3}
\begin{figure}[h]
\begin{center}
\includegraphics[angle=-90,scale=0.30]{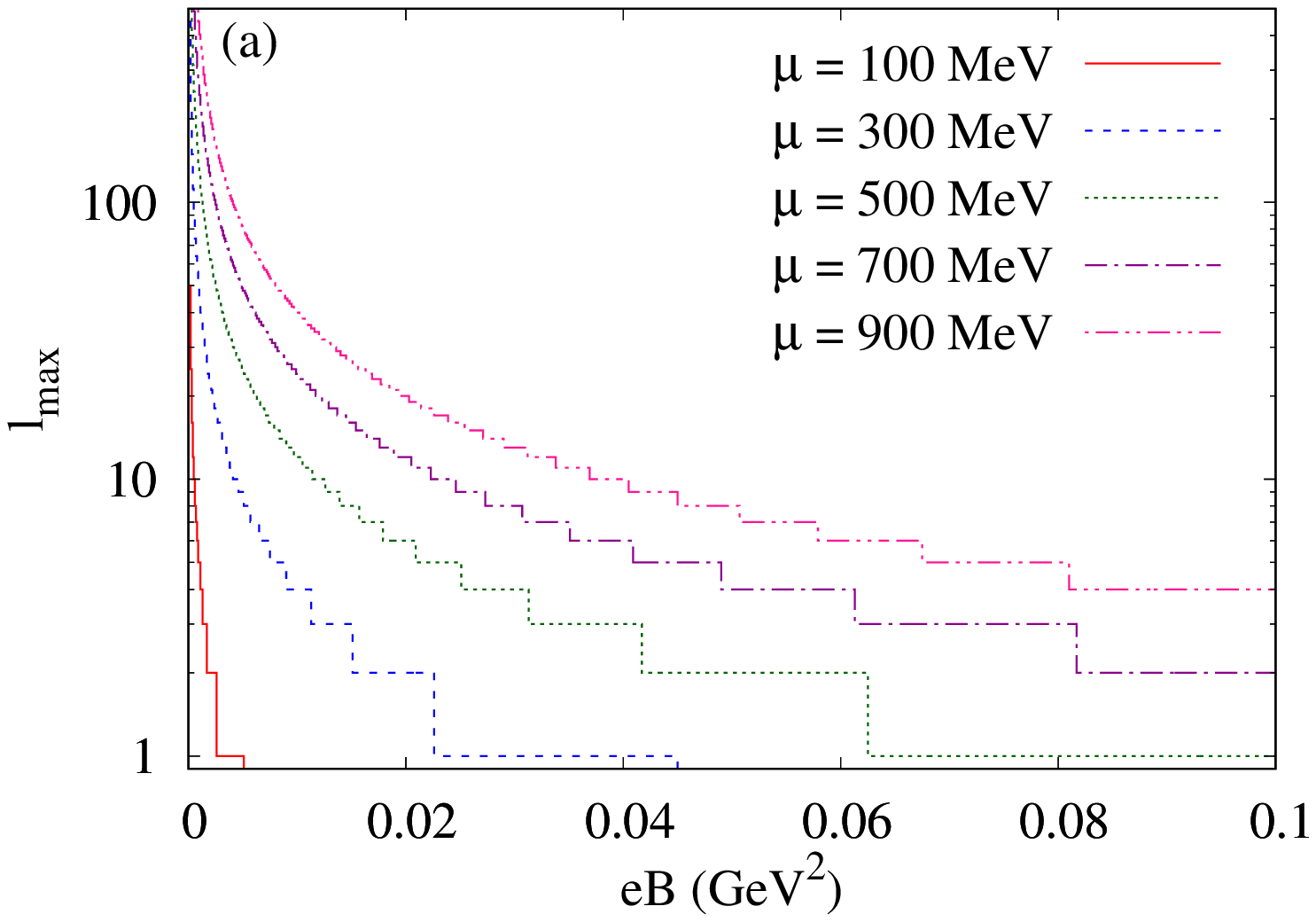} \includegraphics[angle=-90,scale=0.30]{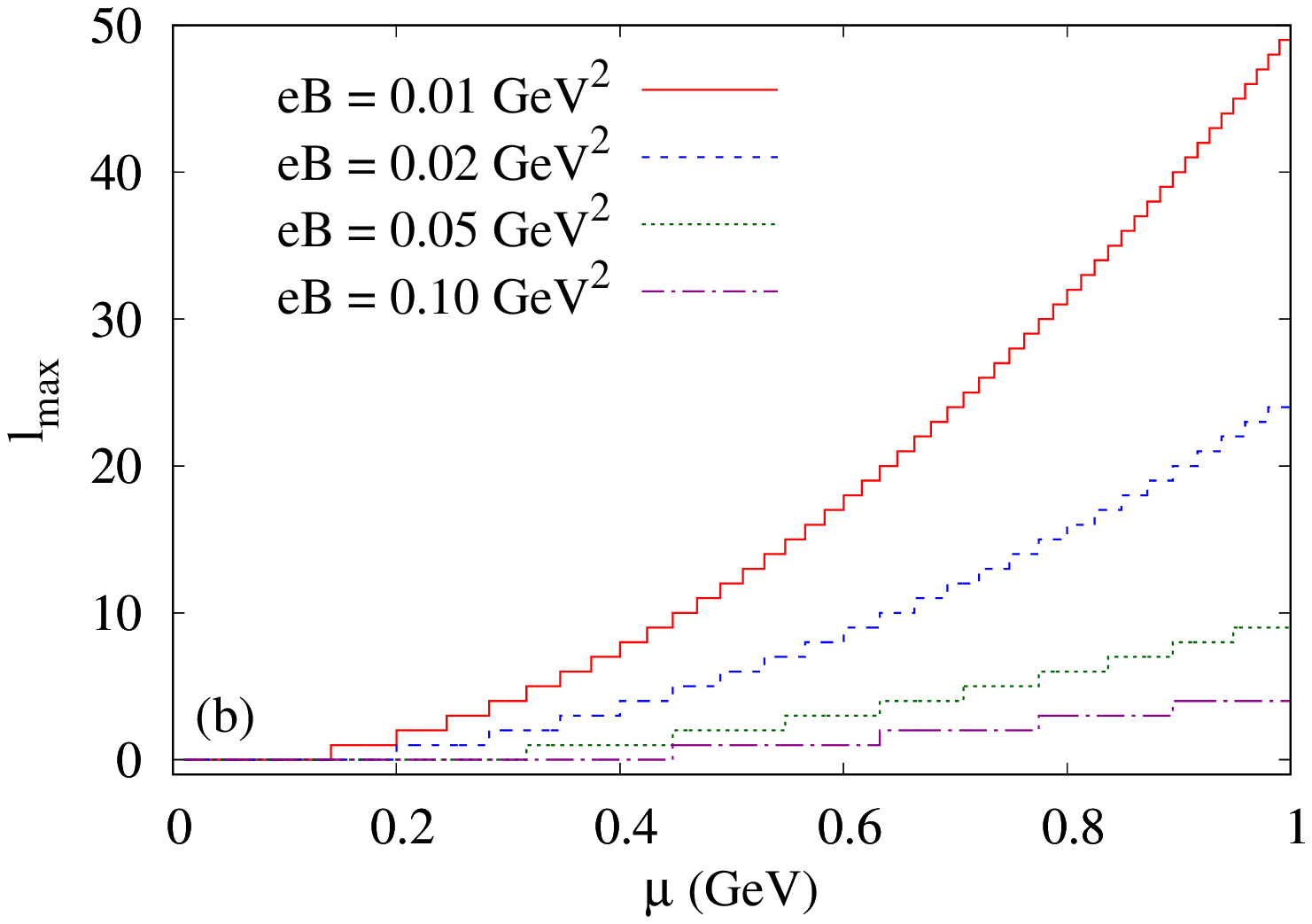}
\end{center}
\caption{(Color Online) The variation of $l_\text{max}$ as a function of (a) magnetic field for different values of chemical potential, and, (b) chemical potential for different values of magnetic field.}
\label{fig.lmax}
\end{figure}
Lets us start our result section first by simple graphical representation of maximum value of Landau level $l_\text{max}$ as a function of $eB$ and $\mu$ in Fig.~\ref{fig.lmax}(a) and (b), which will be very useful to understand our latter results. We have plotted integers values of $l_\text{max}= \left\lfloor \frac{\mu^2-m^2}{2eB} \right\rfloor$ vs $eB$ and $\mu$ in Fig.~\ref{fig.lmax}(a) and (b) for massless fermionic matter at different values of $\mu$ and $eB$ respectively. Here one can see a rough transition from continuous to quantized pattern as we increase $eB$ and decrease $\mu$. It may be considered as a visual transition from classical zone with low $eB$ and/or high $\mu$ to quantum zone with high $eB$ and/or low $\mu$.

Next, our aim to go for numerical discussions of field theoretical expressions of conductivity components, given in Eqs.~\eqref{eq.sigma.pll.3} and \eqref{eq.sigma.perp.3}. To understand the field theoretical contribution in these expressions, let us have a quick recapitulation of the classical and quantum mechanical expressions of conductivity in magnetic field from Ref.~\cite{Dey:2021fbo}. 
In the presence of magnetic field, the electrical conductivity at zero temperature and finite $\mu$ is obtained from RTA formalism $\sigma_{\parallel,\perp}$ as~\cite{Dey:2021fbo}
\begin{eqnarray}
\sigma^\text{RTA}_{\parallel,\perp} = 2 e^2 \int\frac{d^3k}{(2\pi)^3}\tau_c^{\parallel,\perp}
\frac{\vec{k}^2}{3\omega_k^2} \delta(\omega_k -\mu)
=\frac{e^2}{3\pi^2}\frac{\big(\mu^2-m^2\big)^{3/2}}{\mu} \tau_c^{\parallel,\perp} 
\label{RTA}
\end{eqnarray}
where, $\omega_{k}=\sqrt{\vec{k}^2+m^2}$ is the single particle energy and $\tau_c^{\parallel,\perp}$ are the effective relaxation times given by
\begin{eqnarray}
\tau_c^{||} &=& \tau_c, \\
\tau_c^{\perp} &=& \frac{\tau_c}{1 + \frac{\tau_c^2}{\tau_B^2}},
\label{tauc-eff}
\end{eqnarray} 
in which $\tau_c$ is the relaxation time in the absence of magnetic field and $\tau_B=\frac{\mu}{eB}$ is the inverse of cyclotron frequency. It is to be noted that, the actual definition of $\tau_B$ is given by $\tau_B(\omega_k) = \frac{\omega_k}{eB}$ which becomes $\tau_B=\frac{\mu}{eB}$ in Eq.~(\ref{RTA}) due to the presence of the Dirac delta function in the integrand. The RTA expression of $\sigma_{\parallel,\perp}$, which may be called as classical mechanical (CM) expressions, gets modified on the imposition of Landau quantization to the energy-momentum relation of charged particles and we obtain the quantum mechanical (QM) version of $\sigma_{\parallel,\perp}$. Classical to quantum transformation is done by hand as to modify the energy-momentum relation $\omega_{k} = \sqrt{\vec{k}^2 + m^2}\to\omega_{kl} = \sqrt{k_z^2 + m^2 + 2leB}$, where $k_z$ is the momentum along $\hat{\bm{z}}$-direction (in the direction of magnetic field) and $l$ is the Landau level. The modified QM expressions are given by~\cite{Dey:2021fbo}
\begin{eqnarray} 
\sigma_\parallel^\text{QM} &=&
\sum_{l=0}^\infty \alpha_l {e}^2   \frac{eB}{2\pi^2} 
\int\limits^{\infty}_{0} dk_z \frac{k_z^2}{\omega^2_{kl}} \tau_c^{\parallel} \delta(\omega_{kl} -\mu)
=\frac{e^2}{2\pi^2}eB\sum_{l=0}^{l_\text{max}}\alpha_l\frac{\sqrt{\mu^2-2leB-m^2}}{\mu}\tau_c^{\parallel}, \label{qm-pll}\\
\sigma_{\perp}^\text{QM} &=&
\sum_{l=0}^\infty \alpha_l {e}^2   \frac{eB}{2\pi^2} 
\int\limits^{\infty}_{0} dk_z \frac{leB}{\omega^2_{kl}} \tau_c^{\perp} \delta(\omega_{kl} -\mu)
=\frac{e^2}{2\pi^2}eB\sum_{l=1}^{l_\text{max}}\frac{2leB}{\mu\sqrt{\mu^2-2leB-m^2}}\tau_c^{\perp},
\label{qm-perp}
\end{eqnarray} 
where $\alpha_l = (2-\delta_l^0)$ and the summation has been performed upto the highest Landau level $l_\text{max}=\left\lfloor \frac{\mu^2-m^2}{2eB} \right \rfloor $. Similar to the RTA expressions $\sigma_{\parallel,\perp}^\text{RTA}$ and quantum mechanical expressions $\sigma_{\parallel,\perp}^\text{QM}$, we can call the Kubo expressions of Eqs.~\eqref{eq.sigma.pll.3} and \eqref{eq.sigma.perp.3} as $\sigma_{\parallel,\perp}^\text{QFT}$, since the calculations are based on quantum field theory (QFT) at finite $\mu$ and $B$. 

Next, we have generated the curves for conductivity components of RTA, QM and QFT. In all the graphs we have considered the variations of the dimensionless quantity $\sigma/(\tau_c\mu^2)$ with $eB$ and $\mu$. We have also fixed $\tau_c = \Gamma^{-1} = 10$ fm for all the results. Interesting part of our results is the oscillatory behaviour of QM and QFT curves, which can be realized as so-called Shubnikov-de Haas (SdH) effect~\cite{Schubnikow:1930.1,Schubnikow:1930.2,Schubnikow:1930.3,Schubnikow:1930.4,Inagaki:2003ac,Elmfors:1993bm,Elmfors:1993wj} or SdH oscillations. In this effect, at low temperatures and at intense magnetic fields, the electrical conductivity can oscillate. This phenomena is quite well known in condense matter physics. Present work indicates a possibility of this phenomena in compact star environment, where also a strong magnetic field is expected in dense matter.
\begin{figure}[h]
\includegraphics[angle=-90,scale=0.30]{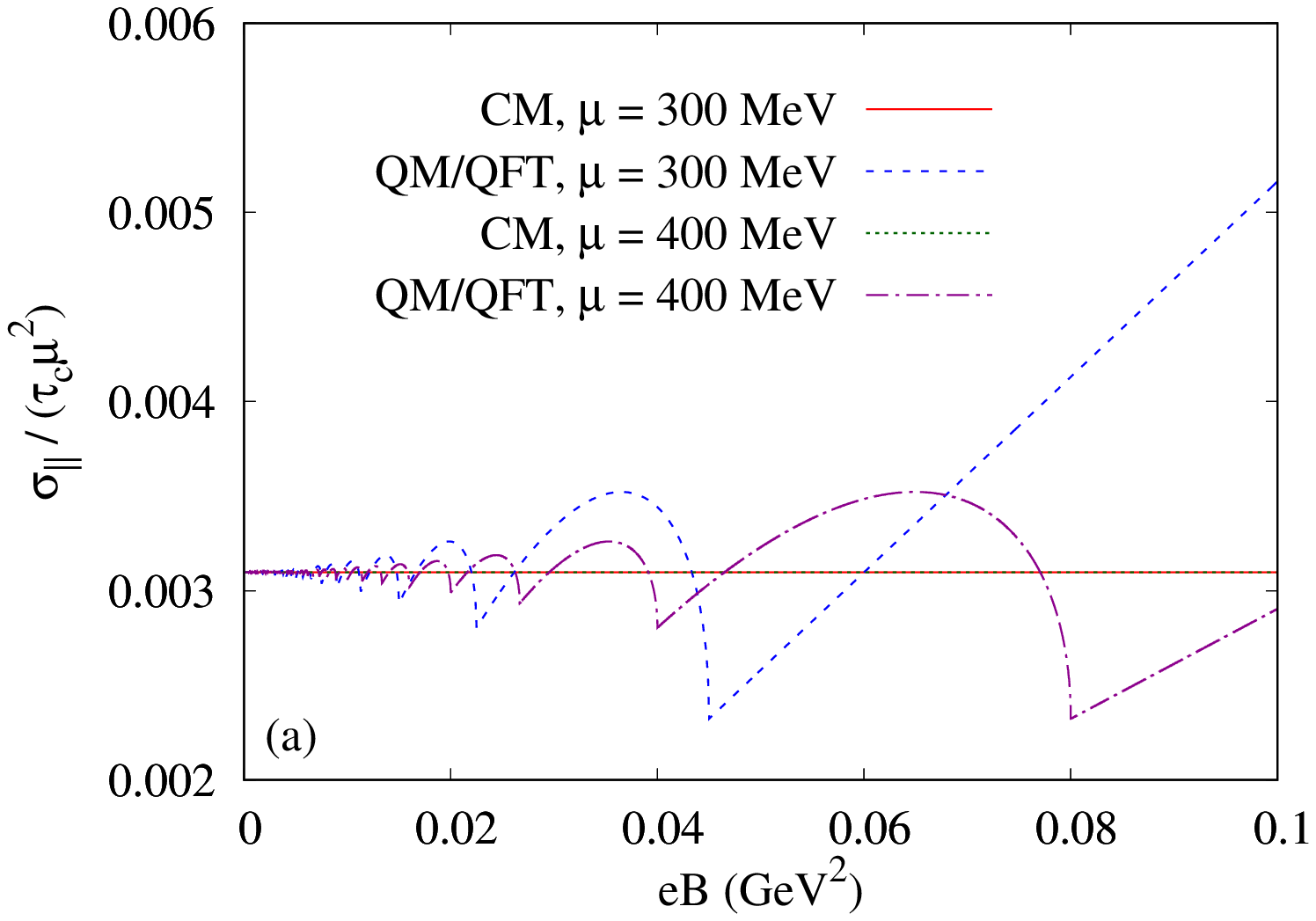} \includegraphics[angle=-90,scale=0.30]{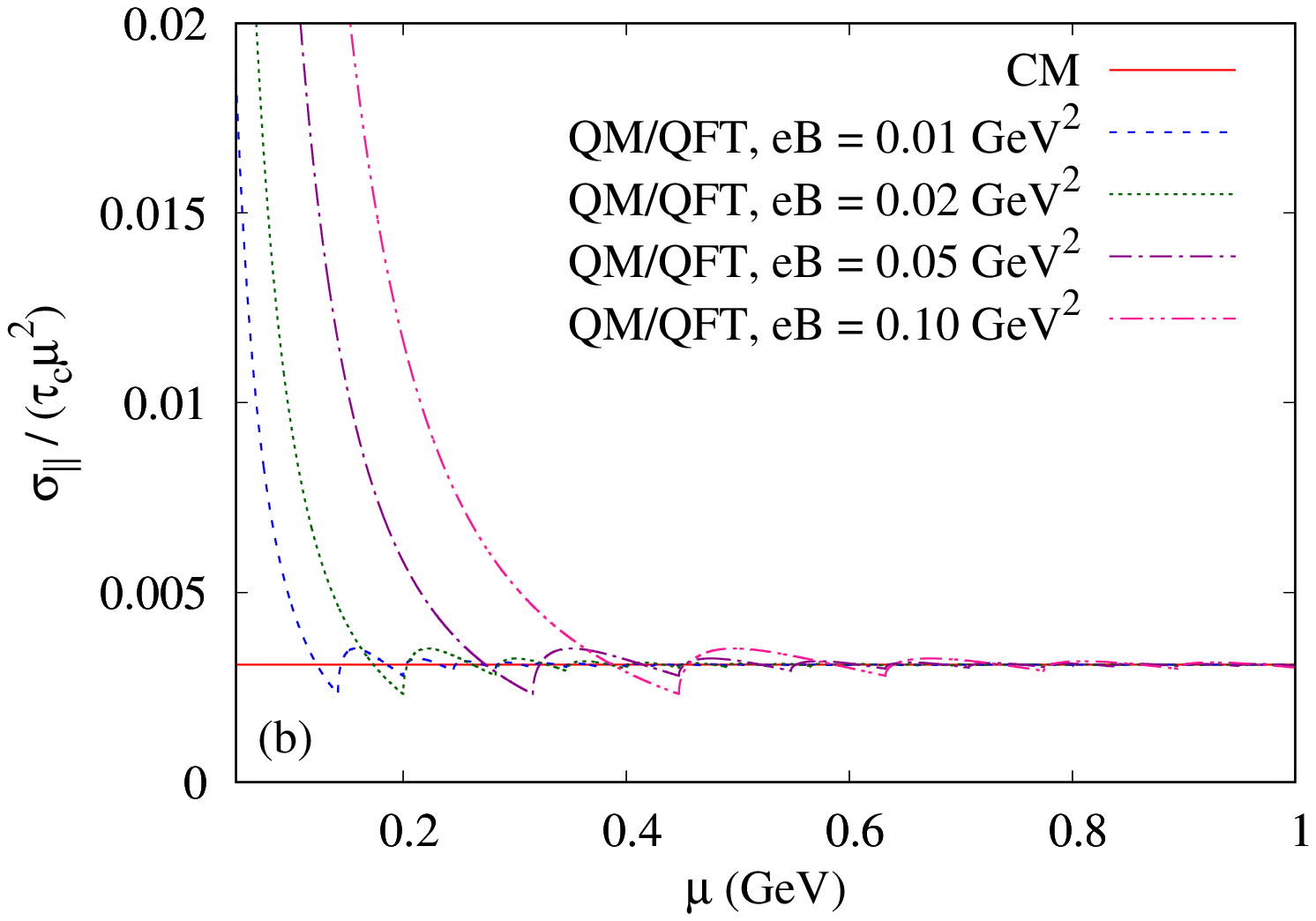}
\caption{The variation of CM and QM/QFT values of $\sigma_\parallel/(\tau_c\mu^2)$ as a function of (a) magnetic field for different values of $\mu$, and, (b) fermion chemical potential for different values of $eB$.}
\label{fig.sigma.pll}
\end{figure}

It has also been seen that, SdH oscillations affect the transport and thermodynamic properties of the material~\cite{Inagaki:2003ac}. It was shown in Ref.~\cite{Inagaki:2003ac}, that the free energy of a compact star exhibits prominent oscillatory modes at low temperatures and vanishing modes at higher temperatures. The results of Ref.~\cite{Inagaki:2003ac} is one of the pioneering works where the oscillations were observed to occur in the thermodynamic properties quark matter. The origin of this effect is purely quantum mechanical. In the forthcoming paragraphs we have explored that quantum effect by comparing RTA, QM and QFT curves, where field theoretical changes are our central attention as a main contribution.

In Figs.~\ref{fig.sigma.pll}(a) and \ref{fig.sigma.pll}(b), we have plotted the variation of $\sigma^{||}/\big(\tau_c\mu^2\big)$ as a function of $eB$ and $\mu$ for CM/RTA, QM and QFT estimations. Comparing Eqs.~\eqref{eq.sigma.pll.3} and \eqref{qm-pll}, we see that the QM expression $\sigma_\parallel^\text{QM}$ exactly matches with the corresponding QFT expression $\sigma_\parallel^\text{QFT}$. However, the QM expressions are obtained by impossing Landau quantization to the CM expression by hand, whereas, the QFT expression are obtained from microscopic calculation in terms of the in-medium spectral function. In Fig.~\ref{fig.sigma.pll}(a), the red colored solid horizontal line is the RTA curve which is found to be independent of magnetic field. This is expected in classical picture as Lorentz force does not do any work in the direction of magnetic field, but this argument does not work for QM or QFT picture. We can notice a  magnetic field dependent parallel conductivity, shown by blue line in Fig.~\ref{fig.sigma.pll}(a), which can not be explained by classical Lorentz force only. Here we see the effect of the Landau quantization of energies which give rise to the SdH effect leading to an oscillatory graph as shown by Fig.~\ref{fig.sigma.pll}(a).

From Fig.~\ref{fig.lmax}(a), we already understood that lesser number of Landau levels will contribute in conduction as the magnetic field will increase. Similarly, for fixed values of $eB$, as we decrease $\mu$, lesser number of Landau levels will contribute in conduction, which is shown in Fig.~\ref{fig.lmax}(b). Hence, the low $\mu$ and large $eB$ domain, where lesser number of Landau levels will contribute, can be consider as quantum domain because the microscopic energy quantization can be revealed in macroscopic quantity such as the conductivity. An oscillatory conductivity in low $\mu$ and high $eB$ zone is observed because of lesser number of Landau levels will contribute in conduction. Reader can understand this fact from Figs.~\ref{fig.sigma.pll}(a) and \ref{fig.sigma.pll}(b). The QM results were also disclosed in Ref.~(\cite{Dey:2021fbo}). So, based on present work and earlier Ref.~(\cite{Dey:2021fbo}), one can expected SdH effect or oscillatory pattern in the  conductivity of dense quark matter, which may exist in the core of NS~\cite{Annala:2019puf}.

On the other hand, one can call the low $eB$ and high $\mu$ domain as classical domain since the CM and QM/QFT curves of Figs.~\ref{fig.sigma.pll}(a) and \ref{fig.sigma.pll}(b) are merged here. Reader might be misguided by seeing the QM/QFT expressions, given in Eqs.~\eqref{eq.sigma.pll.3} and \eqref{qm-pll}, which are proportional to $eB$ and seemed to be zero in the $eB\to 0$ limit but it is not true. For small values of $eB$, $l_{max}$ of Eq.~\eqref{eq.lmax} will be quite larger and it is through larger no of Landau level summation, conductivity will converge towards a finite values instead of being zero. In the limit $eB\to0$, the Landau levels will be infinitesimally close to each other and the contributions from an infinite number of such Landau levels add up to the exact continuum result of CM estimation. In numerical point of view, reaching $eB=0$ for QM/QFT curves are quite difficult as for that case, infinite number of Landau levels has to be summed ideally.

\begin{figure}[h]
\includegraphics[angle=-90,scale=0.30]{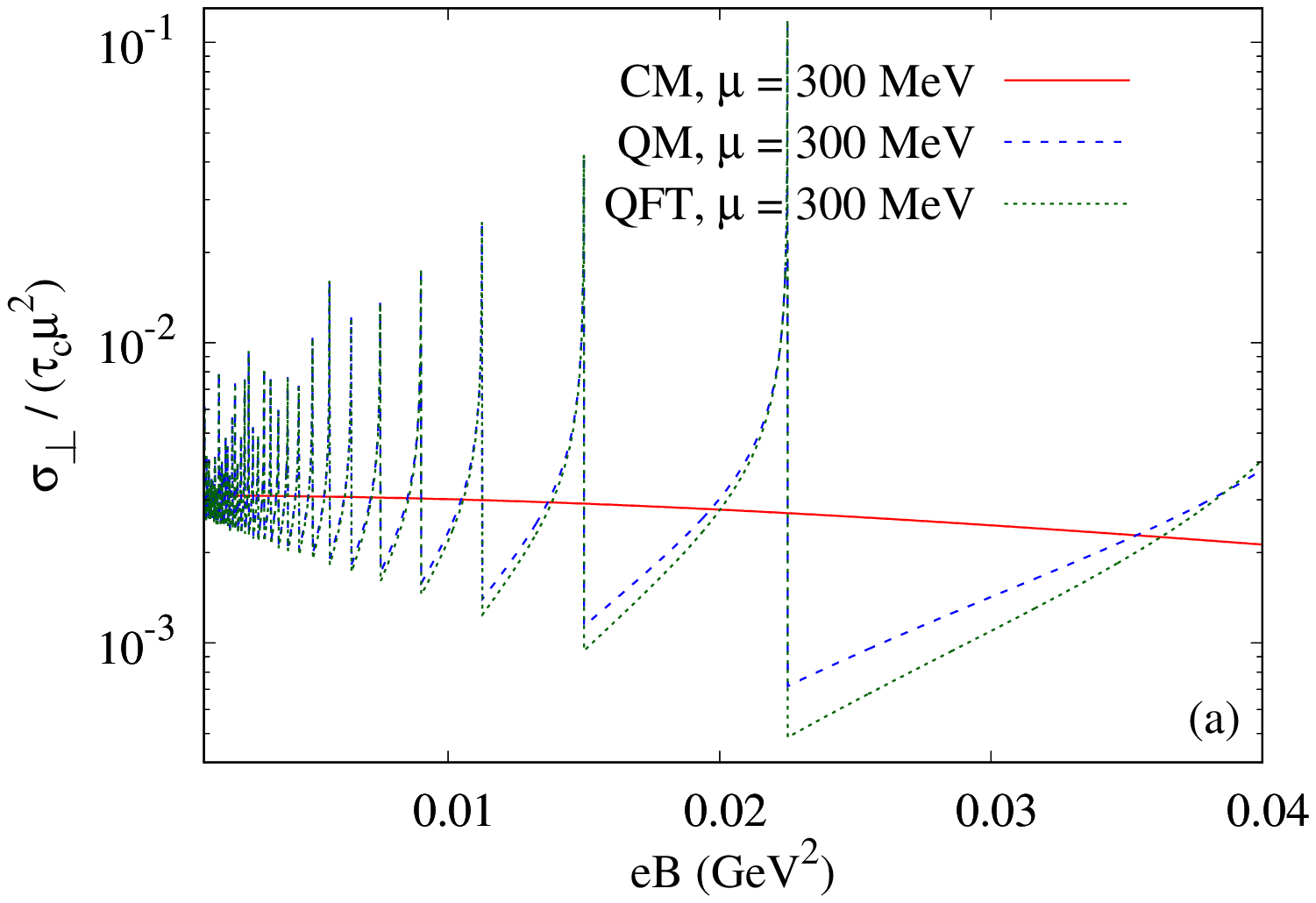} \includegraphics[angle=-90,scale=0.30]{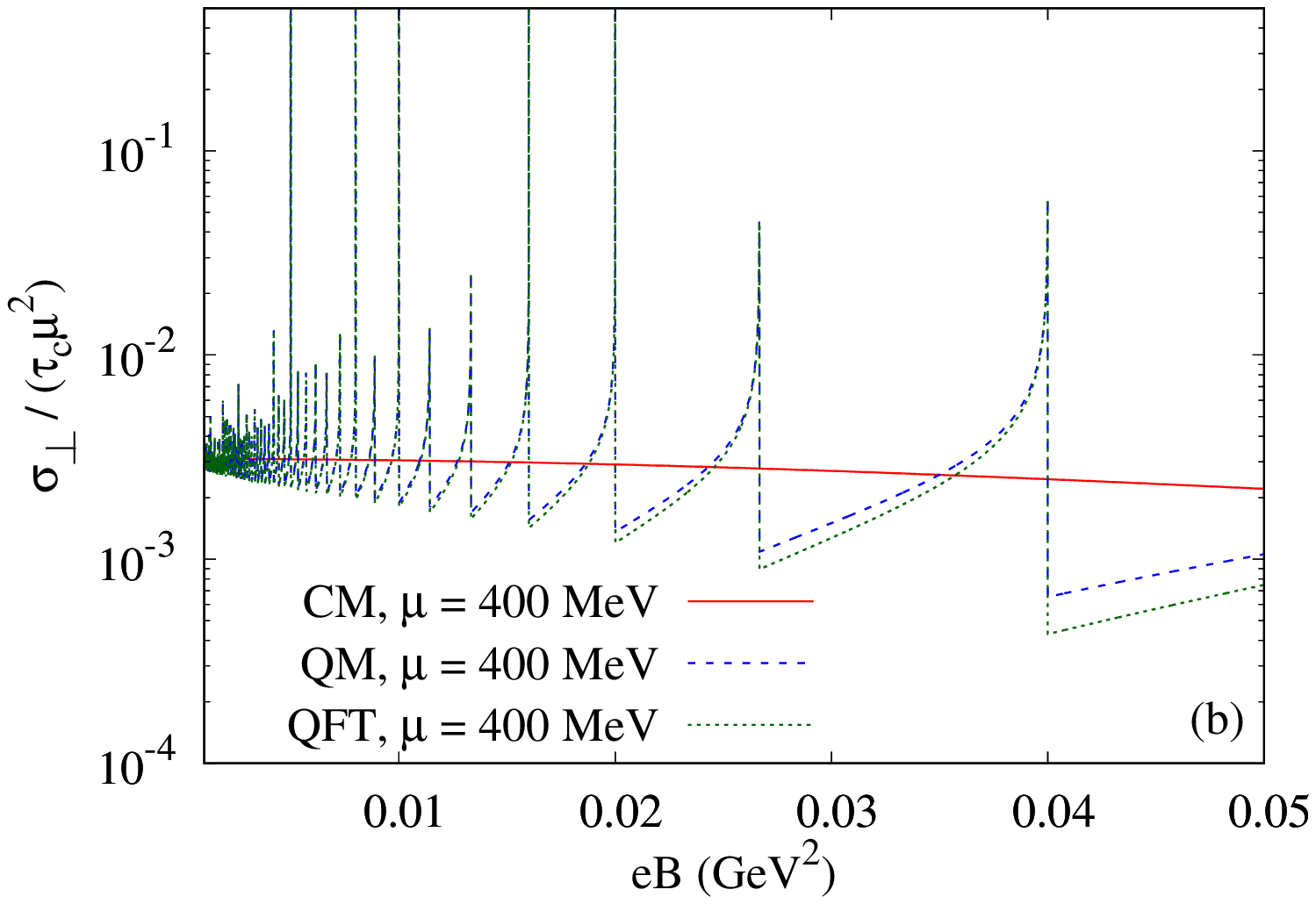} 
\includegraphics[angle=-90,scale=0.30]{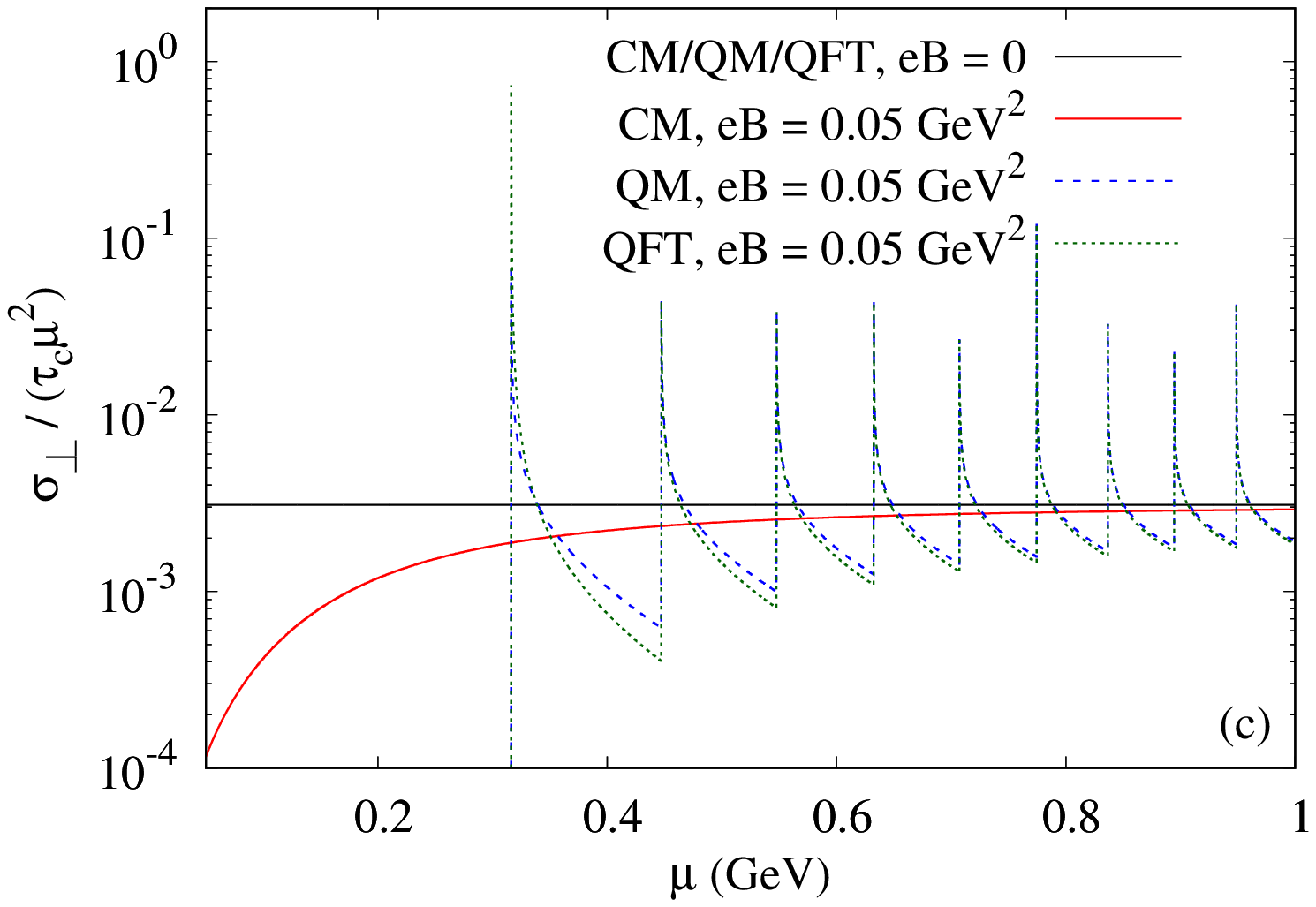} \includegraphics[angle=-90,scale=0.30]{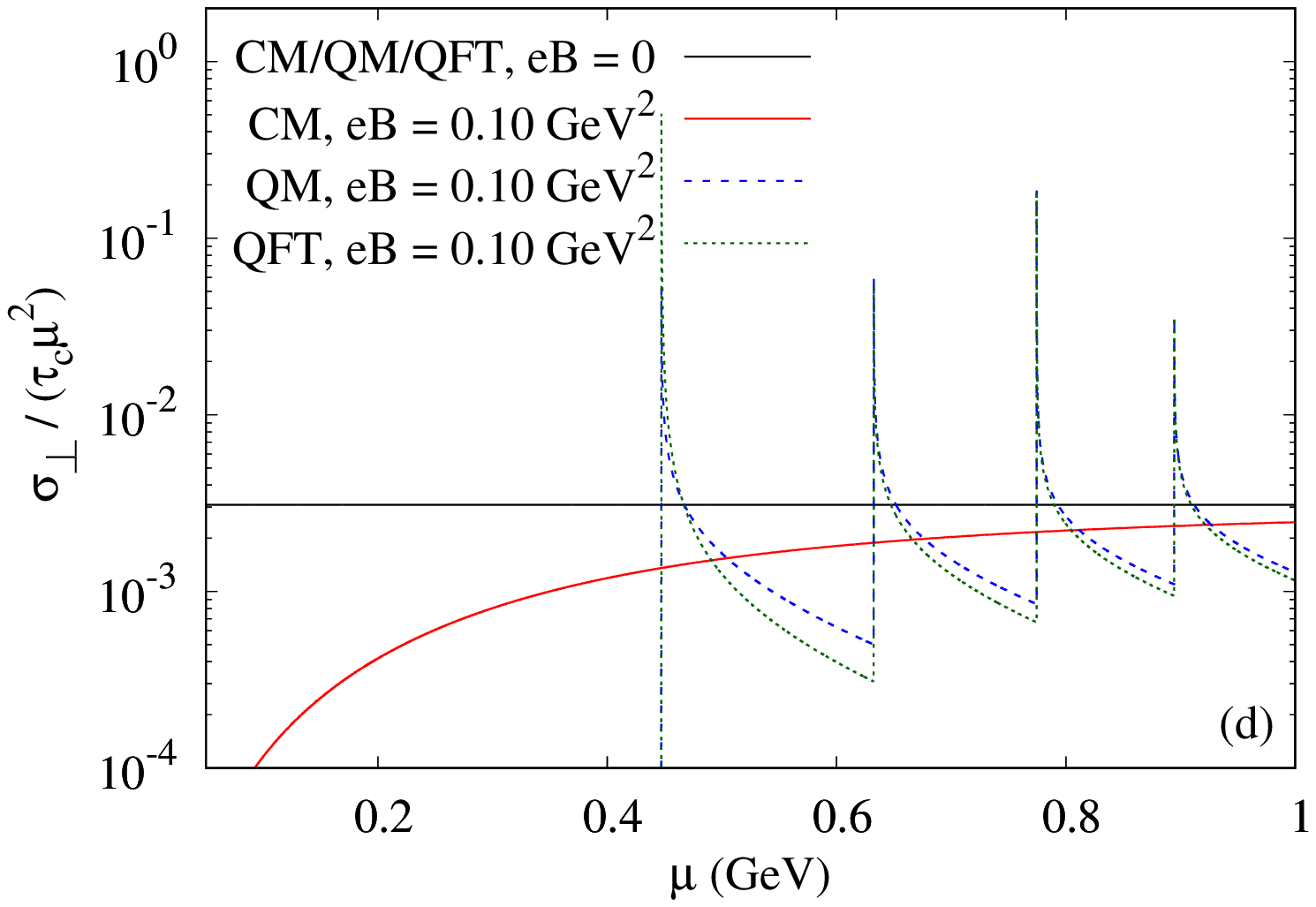}
\caption{The variation of CM, QM and QFT estimation of $\sigma_\perp/(\tau_c\mu^2)$ as a function of (a) magnetic field for $\mu=300$ MeV, (b) magnetic field for $\mu=400$ MeV, (c) fermion chemical potential for $eB=0.05$ GeV$^2$, and, (d) fermion chemical potential for $eB=0.10$ GeV$^2$. The corresponding $eB=0$ graphs are also shown in subfigures (c) and (d) for comparison.}
\label{fig.sigma.per}
\end{figure}

Next in Figs.~\ref{fig.sigma.per}(a)-(d), we have shown the variation of $\sigma^{\perp}/(\tau_c\mu^2)$ as a function of $eB$ and $\mu$ respectively. Unlike to the  CM expression of $\sigma^{\parallel}/(\tau_c\mu^2)$, its perpendicular component has additional $\mu$ and $eB$ dependence due to the factor $1\Big/\SB{1+\FB{\frac{\tau_ceB}{\mu}}^2}$ in Eqs.~\eqref{RTA} and \eqref{tauc-eff}. The factor for small values of $eB$ and large values of $\mu$ become close to one. In Fig.~\ref{fig.sigma.per}(c) and (d), horizontal black solid line indicates CM curve at $eB=0$ and red solid line is perpendicular component of CM curves for two different values of $eB$. They are merging at high $\mu$ but the perpendicular component is suppressed from horizontal line due to the factor. When we generate their QM (dash line) and QFT (dotted line) curves by using Eqs.~\eqref{qm-perp} and \eqref{eq.sigma.perp.3}, we get oscillatory pattern for both cases. Unlike to parallel component case, the quantitative values of QM and QFT expressions for parallel conductivity are different, which can be understood either by comparing Eqs.~\eqref{qm-perp} and \eqref{eq.sigma.perp.3} or by minutely noticing QM (dash line) and QFT (dotted line) curves in Figs.~\ref{fig.sigma.per}(c) and (d). This difference is one of the new finding with respect to earlier Ref.\cite{Dey:2021fbo}, where QM estimations are presented. And also, we are probably first time revealing this new QFT curve in dense sector, whose difference from QM curve is probably connected with a rich field theoretical effect. The peaks are due to the divergent term $\frac{1}{\sqrt{\mu^2 - m^2 -2l_\text{max}B}}$ in the expression of $\sigma^\text{QM}_{\perp}$ and $\sigma^\text{QFT}_{\perp}$. These oscillatory with peak pattern will be faded as we transit from quantum domain (low $\mu$ and high $eB$) to classical domain (high $\mu$ and low $eB$). The merging of CM, QM and QFT curves at high $\mu$ in Fig.~\ref{fig.sigma.per}(c), (d) and low $eB$ in Fig.~\ref{fig.sigma.per}(a), (b) is disclosing that fact.  

At low $\mu$, the QM and QFT values of perpendicular component can be zero, which is connected with the Lowest Landau Level (LLL) approximation. Interestingly, LLL approximation of parallel component will give us non-zero values:
\begin{eqnarray}
\sigma_\parallel^\text{LLL} &=& e^2 \FB{\frac{eB}{2\pi^2}}\frac{\tau_c}{ \mu} \sqrt{\mu^2-m^2}\Theta(\mu-m),~~\text{so that,} \nn \\
\frac{\sigma_{\parallel}^\text{LLL}}{\tau_c\mu^2} &=& \frac{e^2}{2\pi^2} \FB{\frac{eB}{\mu^2}} ~\text{for massless case}~,
\end{eqnarray}
but perpendicular component in LLL approximation become zero, $\sigma_{\perp}^\text{LLL}=0$. In this extreme low $\mu$ with $\mu^2\leq 2eB$ or high $eB$ with $eB\geq \mu^2/2$, which might be considered as extreme quantum domain, there will be no conduction in the perpendicular/transverse direction. It means that 3D anisotropic conduction picture will be transformed to 1D picture with extreme anisotropic conduction. We may find a particular domain, where quark core in NS can reach this LLL or 1D picture. 
\begin{figure}[h]
\includegraphics[angle=-90,scale=0.33]{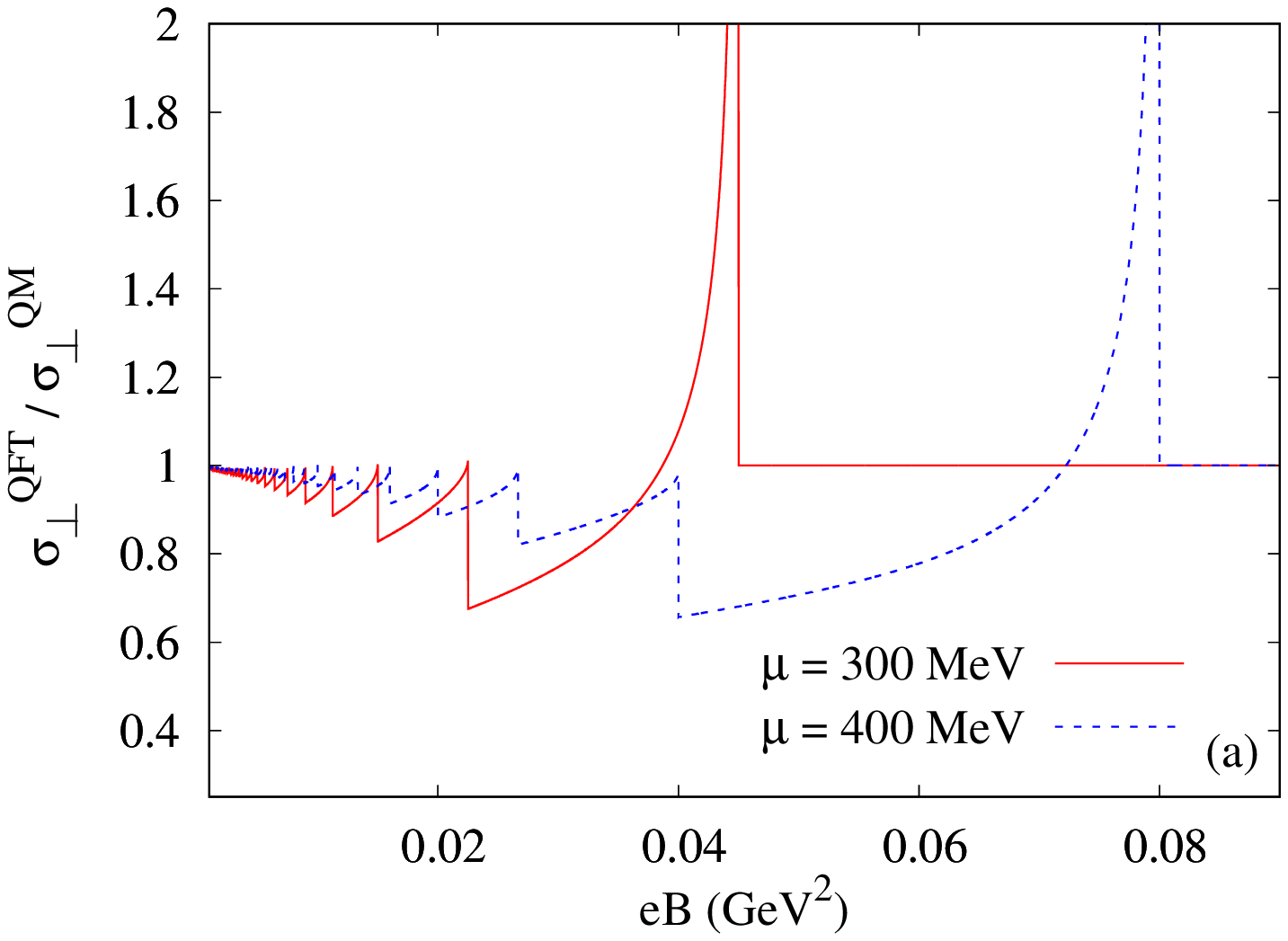} \includegraphics[angle=-90,scale=0.33]{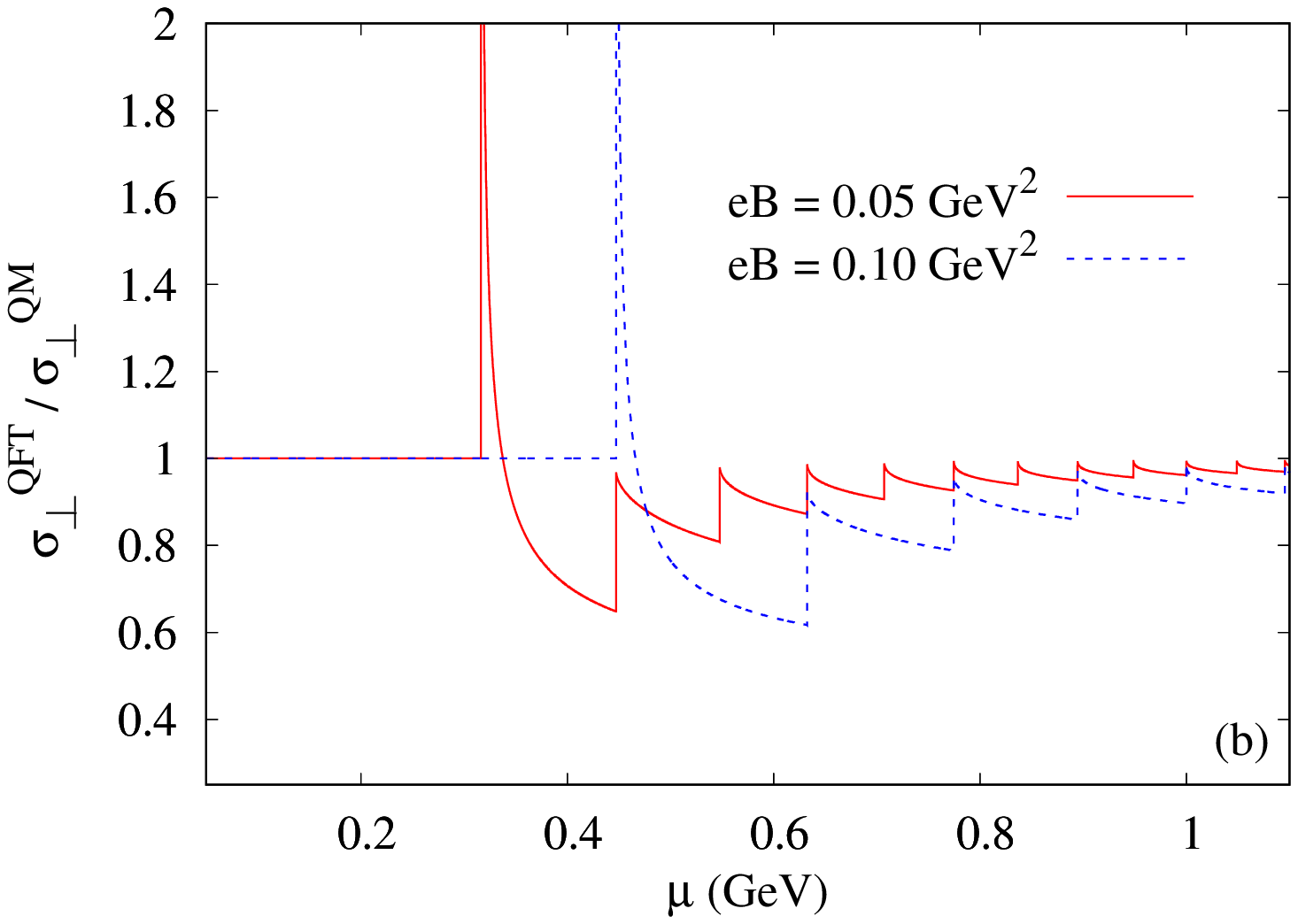} 
\caption{The variation of $\sigma_\perp^\text{QFT}/\sigma_\perp^\text{QM}$ as a function of (a) magnetic field for different values of $\mu$, and, (b) fermion chemical potential for diferent values of $eB$}
\label{fig.sigma.per.ratio}
\end{figure}

Due to the oscillating and spike like nature of $\sigma_{\perp}$ as shown in Fig.~\ref{fig.sigma.per}, it is difficult to feel the actual quantitative difference between the QFT and QM estimations of perpendicular conductivity component. For this, we have depicted the ratio $\sigma_\perp^\text{QFT}/\sigma_\perp^\text{QM}$ as a function of magnetic field and fermion chemical potential in Figs.~\ref{fig.sigma.per.ratio}(a) and (b) respectively. When $eB$ or $\mu$ values changes, the integer value of $l_\text{max}= \left\lfloor \mu^2/(2eB) \right \rfloor  $ remains the same within a particular interval but when $eB$ or $\mu$ values enter into next possible integer value of $l_\text{max}$, then a sudden spike is appeared at the transition values $eB=\mu^2/(2l_\text{max})$ or $\mu=\sqrt{2l_\text{max}eB}$. Skipping those spike values at transition $eB$ or $\mu$ points, we can get a gross profile of the ratio $\sigma_\perp^\text{QFT}/\sigma_\perp^\text{QM}$ along the $eB$ and $\mu$ axes. In quantum domain (i.e. low $\mu$ and high $eB$ regions), the ratio becomes less than unity but it tends to unity in classical domain (i.e high $\mu$ and low $eB$ regions), where QM and QFT both curves merge to CM curve. Hence, this deviation of ratio from unity signifies that QFT estimation carry more enrich quantum effect than simple Landau quantization concept, embedded in QM estimation. Before entering lowest Landau level (where perpendicular transportation freeze), the ratio receive maximum suppression (around $40\%$), which is quite strong.

By comparing CM, QM and QFT expressions of perpendicular components, given in Eqs.~(\ref{RTA}), (\ref{qm-perp}) and (\ref{eq.sigma.perp.3}) respectively, we want to point out another interesting part of our present investigation. Similar to the classical effective relaxation time $\tau_c^\perp=\tau_c \Big/ \FB{1 + \frac{\tau_c^2}{\tau_B^2}}$, we can recognize the QFT based effective relaxation time:
\begin{eqnarray} 
{\widetilde{\tau}}_c^\perp &=& \frac{\Gamma}{\Gamma^2 + \big(\mu - \sqrt{\mu^2 - 2eB}\big)^2} = \tau_c\Big/\FB{1 + \frac{\tau_c^2}{\widetilde{\tau}_B^2}}~,
\end{eqnarray} 
where, 
\begin{eqnarray}
\frac{1}{\widetilde{\tau}_B} = \fb{\mu-\sqrt{\mu^2-2eB}} = \frac{eB}{\mu} \SB{ 1 +\frac{eB}{2\mu^2} +\frac{(eB)^2}{2\mu^4} +\frac{5(eB)^3}{8\mu^6} + \cdots } 
= \frac{1}{\tau_B} \SB{ 1 + \frac{1}{2\mu\tau_B} + \frac{1}{2(\mu\tau_B)^2} + \frac{5}{8(\mu\tau_B)^3} + \cdots }
\end{eqnarray}
One can see that for $\frac{1}{\mu\tau_B}=\frac{eB}{\mu^2}\rightarrow 0$, ${\widetilde{\tau}}_B\rightarrow \tau_B$ as expected. So inverse of cyclotron frequency $\tau_B=\frac{\mu}{eB}$ in classical domain, i.e. small $eB$ and large $\mu$, will transform into 
\begin{eqnarray}
{\widetilde{\tau}}_B &=&\tau_B \Big/ \Big\{1+\frac{1}{2}\Big(\frac{eB}{\mu^2}\Big)
+\frac{1}{2}\Big(\frac{eB}{\mu^2}\Big)^2+\frac{5}{8}\Big(\frac{eB}{\mu^2}\Big)^3+...\Big\}~, 
\label{tB_QFT}
\end{eqnarray}
as we go towards quantum domain, i.e. large $eB$ and small $\mu$. Interestingly we can compare this fact with the transition from non-relativistic (NR) energy momentum relation $E_{NR}=\frac{p^2}{2m}$ to relativistic (R) series-type relation $E_{R}=E_{NR}\Big[1- \frac{1}{4}\Big(\frac{p}{m}\Big)^2+....\Big]$, whose higher order terms become important as we increase the momentum $p$ or velocity gradually. Similarly, as we increase $eB$ and/or decrease $\mu$ or increase $\frac{eB}{\mu^2}$, higher order terms of Eq.~(\ref{tB_QFT}) will be important. So one may get a comparative feelings between this CM to QFT transition for effective cyclotron frequency and the NR to R transition for kinetic energy.   
We are first time addressing this fact, which probably carry very important field theoretical information. 
\begin{figure}[h]
\includegraphics[angle =-90,scale=0.30]{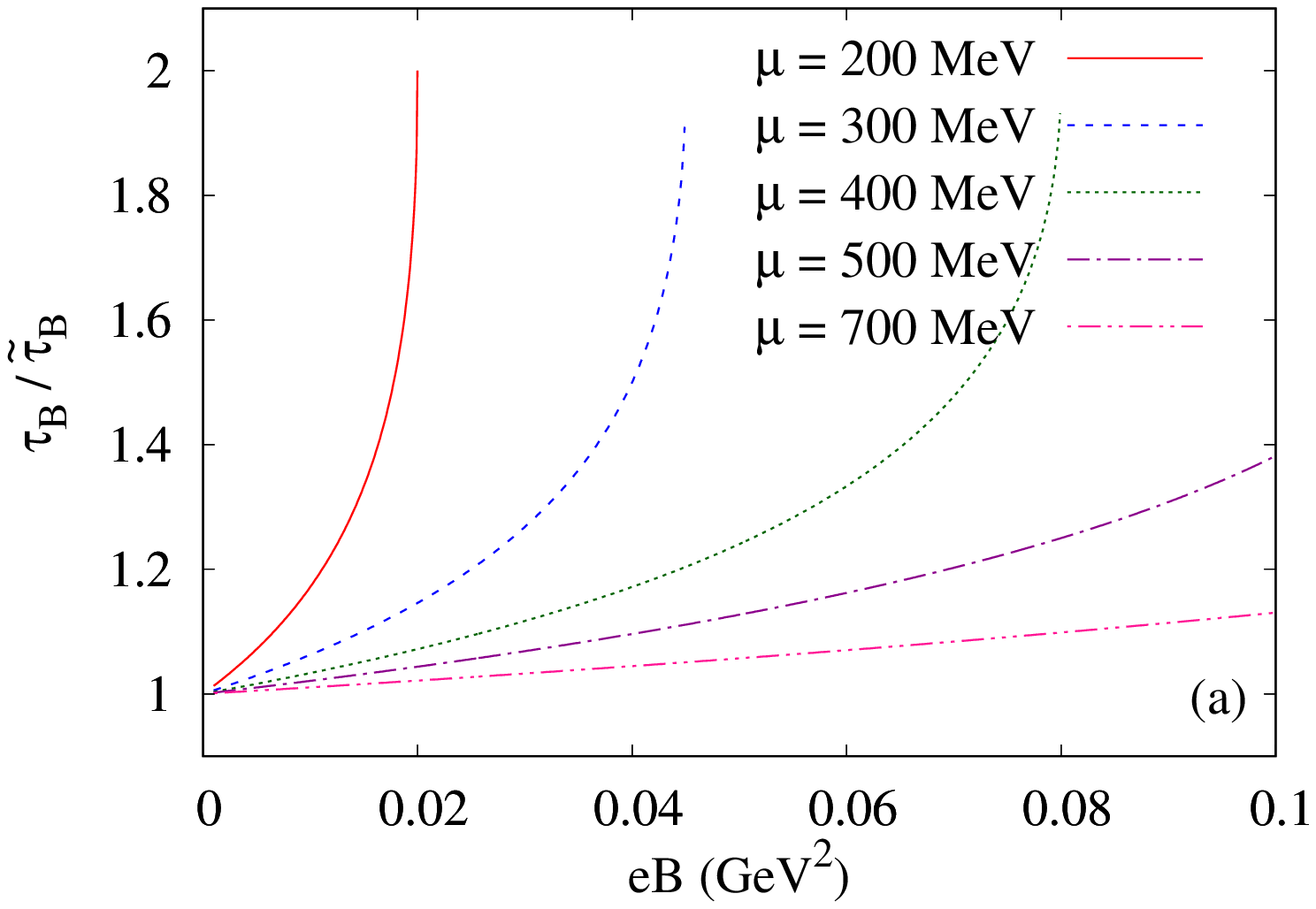}  \includegraphics[angle=-90,scale=0.30]{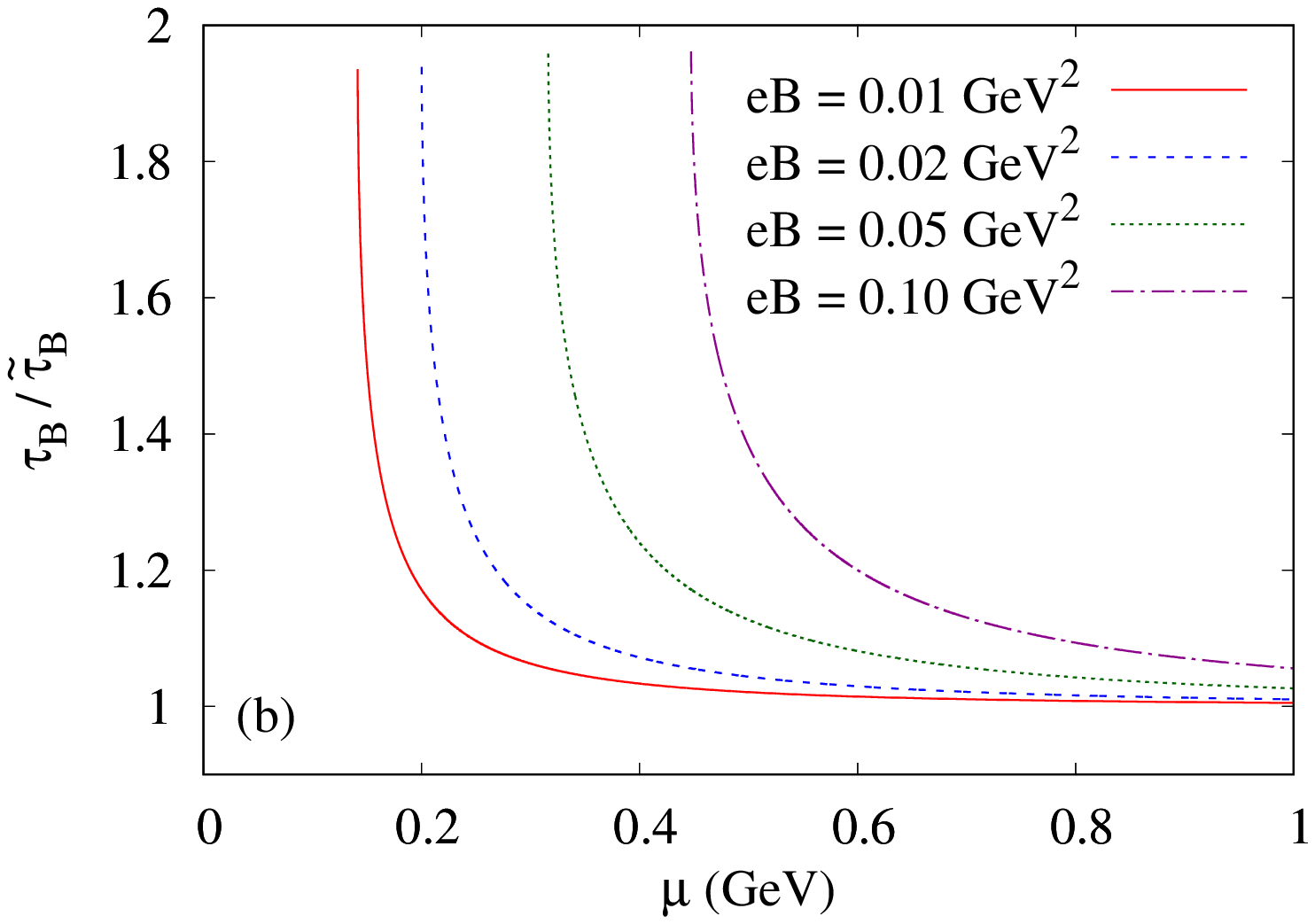}
\caption{The variation of the ratio $\tau_B/\tilde{\tau}_B$ as a function of (a) magnetic field for different values of $\mu$, and, (b) fermion chemical potential for different values of $eB$.}
\label{fig.tau}
\end{figure}

In the quantum mechanical expressions, reader should notice that classical quantity  $\tau_B$ is still present in $\sigma_{\perp}^\text{QM}$, so QM expressions carry semi-classical picture of particle quantization. Actually QM expression is designed by imposing Landau quantization in CM expression. Therefore the momentum integration is only modified but other part remain same as in CM expression. A proper quantized picture is found in the field theoretical expressions $\sigma_{\perp}^\text{QFT}$. This important difference between CM/QM and QFT results manifests itself through an analogy of the anisotropic factor $\dfrac{1}{1 + \frac{\tau_c^2}{\tau_B^2}}$ in CM/QM sector to the factor $\dfrac{\Gamma}{\big(\mu - \sqrt{\mu^2- 2eB}\big)^2 +\Gamma^2}$ in QFT sector. We have taken the ratio of cyclotron time period for CM/QM to that of QFT $\tau_B/{\tilde \tau_B}$ and plotted against $eB$ and $\mu$ in Figs.~\ref{fig.tau}(a) and \ref{fig.tau}(b) respectively, where a clear deviation from one is noticed for quantum domain i.e. high $eB$ and low $\mu$ regions. We see that cyclotron time period for CM is larger than that of QFT in that domain. It means that QFT push the system with larger cyclotron frequency and lesser cyclotron time period, for which anisotropic conduction will also be increased.

%~~~~~~~~~~~~~~~~~~~~~~~~~~~~~~~~~~~~~~~~~~~~~~~~~~~~~~~~~~~~~~~~~~~~~~~~~~~~~~~~~~~~~~~
\section{SUMMARY AND CONCLUSIONS} \label{sec4}
Present work have explored field theoretical structure of electrical conductivity of degenerate relativistic fermionic matter in presence of magnetic field. With the help of Schwinger's proper time formalism, two point function of current-current correlator in one-loop diagram level is obtained for the degenerate fermionic medium. Owing to the Kubo relation, conductivity tensor is realized as zero momentum limit of the current-current correlator, whose one-loop level diagram carry two propagators with two different Landau level sumation. Due to orthogonal properties of Laggure polynomials, conduction along perpendicular direction selects $\pm1$ differences of Landau levels of propagators, while parallel conduction selects propagators with same Landau levels. This fact is well established in our earlier work~\cite{Satapathy:2021cjp} for finite temperature calculation, which is modified in present work for finite density picture, relevant for compact star environment. 

After going through a rigorous calculations, based on real-time formalism thermal field theory, we get at the end very simple algebric relations with Landau level summation. Based on relaxation time approximation (RTA) method, Ref.~\cite{Dey:2021fbo} has obtained similar kind of algebric expressions for degenerate fermionic matter, which is called here as classical mechanical (CM) expressions, and its Landau quantization extension is called as quantum mechanical (QM) expressions for distinguishing them from our quantum field theoretical (QFT) expressions, addressed in present work. Background of CM and QM expressions are the RTA based kinetic theory approach but the background of QFT expression is the Kubo approach. Their background methods are completely different. However, the interesting news, revealed by present work, are that QFT expression is exactly same with QM expression for parallel conductivity component but they are different for perpendicular component. It is non-zero cyclotron frequency, which is entered into the CM expression of perpendicular component of electrical conductivity and responsible for reducing its conduction along perpendicular direction with respect to its parallel component. In QFT expression of perpendicular conductivity, we get a completely new expression of cyclotron frequency, which become quite larger than its classical values in the quantum domain. Low chemical potential and high magnetic field can be considered as quantum domain, where parallel and perpendicular conductivity components get oscillating pattern like Shubnikov-de Haas (SdH) oscillation, which is well known phenomenon in condense matter physics. Present work anticipates a possibility of SdH or quantized pattern of conductivity tensor in compact star environment, which may demand more future works, related with this particular topic from theoretical and phenomenological sides of astro-physics sector.

At the end, let us locate our limitation of present calculation. We are confined here within one-loop calculation but for real system, having interaction Lagrangian density, infinite order ladder diagrams~\cite{Jeon:1994if,Fernandez-Fraile:2005bew} have to considered as they all may contribute in same order of magnitude. Such resummation of the ladder diagrams in presence of external magnetic field in the calculation of longitudinal conductivity of quark matter has been performed using LLL approximation in Ref.~\cite{Hattori:2016cnt} and including all the Landau levels in Ref.~\cite{Fukushima:2019ugr}. At present, there are no other calculations for the transverse conductivity, done in QM and QFT approaches, which can support our findings or bring more clear picture on the issues like whether QFT is more mature expressions than QM or not; and whether difference between QFT and QM is sensible or not. Finally, we also note that, in the Kubo approach at non-zero magnetic field, we have introduced the finite thermal width $\Gamma$ as a parameter (by hand), instead of calculating it from interaction Lagrangian. Depending upon the systems, we have to consider corresponding interaction Lagrangian density and have to calculate $\Gamma$ from it. These all are future scopes for extension of present work.
%
%~~~~~~~~~~~~~~~~~~~~~~~~~~~~~~~~~~~~~~~~~~~~~~~~~~~~~~~~~~~~~~~~~~~~~~~~~~~~
\section*{ACKNOWLEDGEMENTS} 
Sarthak Satpathy thanks Amaresh Jaiswal for getting short visit platform in NISER, Bhubaneswar, funded from DST-INSPIRE faculty with Grants No. DST/INSPIRE/04/2017/000038. Snigdha Ghosh is funded by the Department of Higher Education, Government of West Bengal, India. SS thanks Guruprasad Kadam, Jayanta Dey and Hemant Sharma for valuable discussions. Snigdha Ghosh acknowledges Prof. Sourav Sarkar for immense encouragement and support in all respect.

%~~~~~~~~~~~~~~~~~~~~~~~~~~~~~~~~~~~~~~~~~~~~~~~~~~~~~~~~~~~~~~~~~~~~~~~~~~~~~~~~~~~~~~~~~~~~~~~~~~~~~~~~~~~
\appendix
\section{CALCULATION OF THE TWO-POINT CORRELATION FUNCTION} \label{appendix.A}
In this appendix, we will calculate the two-point vector current correlation function $\Ensembleaverage{\mcTc J^{\mu}(x)J^{\nu}(y)}_{11}^B$ in the presence of external magnetic field. We have from Eq.~\eqref{eq.current} 
\begin{eqnarray}
\ensembleaverage{\mathcal{T}_c J^{\mu}(x)J^{\nu}(y)}_{11}^B  = 
e^2\ensembleaverage{\mathcal{T}_c\psibar(x)\gamma^{\mu}\psi(x)\psibar(y)\gamma^{\nu}\psi(y)}_{11}^B. \label{eq.A1.1}
\end{eqnarray}
Applying Wick's theorem on the RHS of Eq.~\eqref{eq.A1.1} yields
\begin{eqnarray}
\ensembleaverage{\mathcal{T}_c J^{\mu}(x)J^{\nu}(y)}_{11}^B &=& 
e^2 \wick[offset=1.2em]{ \ensembleaverage{\mcTc \c2\psibar(x)\gamma^\mu \c1 \psi(x) \c1 \psibar(y)\gamma^\nu \c2 \psi(y)}^B_{11}} \nn \\
&=& -e^2\Tr\SB{\gamma^{\mu}S^B_{11}(x,y)\gamma^{\nu}S^B_{11}(y,x)}
\label{eq.A1.wick}
\end{eqnarray}
where, $S^B_{11}(x,y) = \wick[offset=1.2em]{\ensembleaverage{\mcTc \c\psi(x) \c\psibar(y) }^B_{11}}$ denotes the 11-component of the coordinate space thermo-dense-magnetic Dirac propagator in RTF. It is to be noted that, Eq.~\eqref{eq.A1.wick} remains valid even if the fermion field $\psi$ is a multiplet, in which case, the traces will have to be taken over all the spaces belonging to the multiplet in addition to the Dirac space. The magnetized Dirac propagator $S^B_{11}(x,y) = \Phi(x,y) S^B_{11}(x-y)$ is not translationally invariant due to the the gauge dependent phase factor $\Phi(x,y)$ (which explicitly breaks the translational invariance), however it can partially be Fourier transformed to the momentum space as
\begin{eqnarray}
S^B_{11}(x,y) = \Phi(x,y)\int\!\!\!\frac{d^4p}{(2\pi)^4}e^{-ip\cdot(x-y)}\FB{-iS^B_{11}(p)}
\label{eq.A1.S11B.xy}
\end{eqnarray}
where, 
$S^B_{11}(p)$ is the 11-component of the momentum space free thermo-dense-magnetic Dirac propagator in RTF whose explicit form reads~\cite{Ayala:2003pv,Schwinger:1951nm,Mukherjee:2017dls}
\begin{eqnarray}
S_{11}^B(p) = \sum_{l=0}^{\infty}(-1)^le^{-\alpha_p}\mathscr{D}_l(p)D_{11}(p_\parallel,m_l)~.
\label{eq.A1.S11B.p.1}
\end{eqnarray}
In the above equation, $l$ denotes the Landau level index, $\alpha_p = -\frac{p_{\perp}^2}{eB} \geq 0$, $m_l = \sqrt{m^2 + 2leB}$ is the 
``\textit{Landau level dependent effective mass}'', $\scrD_l(p)$ contains the complicated Dirac structure involving the Laguerre polynomials as 
\begin{eqnarray}
\scrD_l(p) = \FB{\cancel{p}_\parallel+m}
\TB{\FB{   \mathds{1}+i\gamma^1\gamma^2}L_l(2\alpha_p) - \FB{\mathds{1}-i\gamma^1\gamma^2} L_{l-1}(2\alpha_p)}
- 4\cancel{p}_\perp L^1_{l-1}(2\alpha_p)~,
\end{eqnarray}
with the convention $L_{-1}(z) = L_{-1}^1(z) = 0$, and, 
$D_{11}(p,m)$ is given by
\begin{eqnarray}
D_{11}(p,m) = \TB{\frac{-1}{p^2-m^2+i\epsilon} - \xi(p.u)2\pi i\delta(p^2-m^2)}
\label{eq.A1.D11}
\end{eqnarray} 
in which, $\xi(x)=\Theta(x)f_+(x)+\Theta(-x)f_-(-x)$, $f_\pm(x)=\TB{e^{(x\mp \mu )/T}+1}^{-1}$ are the Fermi-Dirac thermal distribution functions, and, $u^\mu$ is the medium four-velocity. In the Local Rest Frame (LRF) of the medium $u^\mu_\text{LRF}\equiv(1,\vec{0})$.

Substituting Eq.~\eqref{eq.A1.S11B.xy} into Eq.~\eqref{eq.A1.wick}, we get
\begin{eqnarray}
\ensembleaverage{\mcTc J^\mu(x)J^\nu(y)}^B_{11}  &=&
e^2 \Phi(x,y) \Phi(y,x) \int\!\!\!\int\!\!\!\frac{d^4p}{(2\pi)^4}\frac{d^4k}{(2\pi)^4}e^{-i(x-y)\cdot (p-k)}
\Tr\SB{\gamma^{\mu}S^B_{11}(p)\gamma^{\nu}S^B_{11}(k)} 
\label{eq.A1.corr.1}
\end{eqnarray}
Again substituting Eq.~\eqref{eq.A1.S11B.p.1} into Eq.~\eqref{eq.A1.corr.1} and using the fact that $\Phi(x,y) \Phi(y,x) = 1$, we get after bit simplifications
\begin{eqnarray}
\ensembleaverage{\mcTc J^\mu(x)J^\nu(y)}^B_{11} &=& 
-\sum_{l=0}^{\infty} \sum_{n=0}^{\infty}\int\!\!\!\int\!\!\!\frac{d^4p}{(2\pi)^4}\frac{d^4k}{(2\pi)^4} e^{-i(x-y)\cdot(p-k)}
D_{11}(\ppll,m_n)D_{11}(\kpll,m_l) 
\mcN_{ln}^\munu(k,p)
\label{eq.A1.corr.2}
\end{eqnarray}
where, 
\begin{eqnarray}
\mathcal{N}^{\mu\nu}_{ln}(k,p) = -e^2(-1)^{l+n}e^{-\alpha_k-\alpha_p} \Tr \SB{\gamma^\mu \scrD_n(p) \gamma^\nu \scrD_l(k)}.
\label{eq.A1.N.1}
\end{eqnarray}

In the calculation of the electrical conductivity, we actually require the expressions of $\mathcal{N}_{ln}^{\mu\nu}(k,k)$ 
and $\mathcal{\widetilde{N}}_{ln}^{\mu\nu}(\kpll) = {\displaystyle \int}\!\!\dfrac{d^2\kper}{(2\pi)^2}\mathcal{N}_{ln}^{\mu\nu}(k,k)$. 
From Eq.~\eqref{eq.A1.N.1}, we get after evaluating the traces over Dirac matrices  
\begin{eqnarray}
\mathcal{N}^{\mu\nu}_{ln}(k,k) &=& -8e^2\Big[8\FB{2k_{\perp}^{\mu}k_{\perp}^{\nu} - k_{\perp}^2g^{\mu\nu}}\mathcal{B}_{ln}(\kper^2)  + \SB{ 2k_\parallel^{\mu}k_\parallel^{\nu}-\gpll^{\mu\nu}(k_\parallel^2-m^2)}\mathcal{C}_{ln}(\kper^2)\nn \\
&& + g_{\perp}^{\mu\nu}\FB{k_\parallel^2-m^2}\mathcal{D}_{ln}(\kper^2)+ 2\FB{k_\parallel^{\nu}k_{\perp}^{\mu} + k_\parallel^{\mu}k_{\perp}^{\nu}}\mathcal{E}_{ln}(\kper^2)\Big].
\label{eq.A1.N.2}
\end{eqnarray}
where, 
\begin{eqnarray}
\mathcal{B}_{ln}(\kper^2) &=& (-1)^{l+n}e^{-2\alpha_k} L^1_{l-1}(2\alpha_k) L^1_{n-1}(2\alpha_k) ~,
\label{eq.Bnl}\\
\mathcal{C}_{ln}(\kper^2) &=& (-1)^{l+n}e^{-2\alpha_k} 
\SB{L_{l-1}(2\alpha_k) L_{n-1}(2\alpha_k) + L_{l}(2\alpha_k) L_{n}(2\alpha_k)} \label{eq.Cnl}~, \\
\mathcal{D}_{ln}(\kper^2) &=& (-1)^{l+n}e^{-2\alpha_k} 
\SB{L_{l}(2\alpha_k) L_{n-1}(2\alpha_k) + L_{l-1}(2\alpha_k) L_{n}(2\alpha_k)} \label{eq.Dnl}~, \\
\mathcal{E}_{ln}(\kper^2) &=& (-1)^{l+n}e^{-2\alpha_k} 
\SB{L_{l-1}(2\alpha_k) L^1_{n-1}(2\alpha_k) - L_{l}(2\alpha_k) L^1_{n-1}(2\alpha_k) \right. \nn \\ && \left.
+~ L^1_{l-1}(2\alpha_k) L_{n-1}(2\alpha_k)- L^1_{l-1}(2\alpha_k) L_{n}(2\alpha_k)}. \label{eq.Enl}
\end{eqnarray}

Using Eq.~\eqref{eq.A1.N.2}, we obtain
\begin{eqnarray}
\mathcal{\widetilde{N}}_{ln}^{\mu\nu}(\kpll) =  \int\!\!\dfrac{d^2\kper}{(2\pi)^2}\mathcal{N}_{ln}^{\mu\nu}(k,k) = 
8e^2\Big[
8\mathcal{B}_{ln}^{(2)}\gpll^\munu
- \mathcal{C}_{ln}^{(0)}\SB{ 2k_\parallel^{\mu}k_\parallel^{\nu}-\gpll^{\mu\nu}(k_\parallel^2-m^2)} - \mathcal{D}_{ln}^{(0)}\FB{k_\parallel^2-m^2}g_{\perp}^{\mu\nu}
\Big].
\label{eq.A1.Ntil.1}
\end{eqnarray}
where\begin{eqnarray}
\mathcal{B}_{ln}^{(j)} &=& \int\!\! \frac{d^2\kper}{(2\pi)^2}\mathcal{B}_{ln}(\kper^2) \FB{\kper^2}^{j/2}, \label{eq.A1.Bln.j} \\
\mathcal{C}_{ln}^{(j)} &=& \int\!\! \frac{d^2\kper}{(2\pi)^2}\mathcal{C}_{ln}(\kper^2) \FB{\kper^2}^{j/2}, \label{eq.A1.Cln.j} \\
\mathcal{D}_{ln}^{(j)} &=& \int\!\! \frac{d^2\kper}{(2\pi)^2}\mathcal{D}_{ln}(\kper^2) \FB{\kper^2}^{j/2}. \label{eq.A1.Dln.j} 
\end{eqnarray}

Exploiting the orthogonality of the Laguerre polynomials present in the functions $\mathcal{B}_{ln}(\kper^2)$, $\mathcal{C}_{ln}(\kper^2)$, and,  $\mathcal{D}_{ln}(\kper^2)$, the $d^2\kper$ integrals of Eqs.~\eqref{eq.A1.Bln.j}-\eqref{eq.A1.Dln.j} are performed and the analytic expressions of the quantities $\mathcal{B}^{(j)}_{ln}$, $\mathcal{C}^{(j)}_{ln}$, and,  $\mathcal{D}^{(j)}_{ln}$ are provided in Appendix~\ref{appendix.B}. 
On substituting Eqs.~\eqref{eq.A2.B2}-\eqref{eq.A2.D0} into Eq.~\eqref{eq.A1.Ntil.1}, we finally arrive at
\begin{eqnarray}
\mathcal{\widetilde{N}}_{ln}^{\mu\nu}(\kpll) &=&  \int\!\!\dfrac{d^2\kper}{(2\pi)^2}\mathcal{N}_{ln}^{\mu\nu}(k,k) \nn \\
&=& e^2  \frac{eB}{\pi} \Big[
-4eBn\delta_{l-1}^{n-1} \gpll^\munu
- \FB{\delta_{l}^{n}+\delta_{l-1}^{n-1}}\SB{ 2k_\parallel^{\mu}k_\parallel^{\nu}-\gpll^{\mu\nu}(k_\parallel^2-m^2)} 
+ \FB{\delta_{l}^{n-1} + \delta_{l-1}^{n}}\FB{k_\parallel^2-m^2}g_{\perp}^{\mu\nu}
\Big].
\label{eq.A1.Ntil.2}
\end{eqnarray}

%~~~~~~~~~~~~~~~~~~~~~~~~~~~~~~~~~~~~~~~~~~~~~~~~~~~~~~~~~~~~~~~~~~~~~~~~~~~~~~~~~~~~~~~~~~~~~~~~~~~~~

\section{ANALYTIC EXPRESSIONS OF $\mathcal{B}_{ln}^{(j)}$, $\mathcal{C}_{ln}^{(j)}$, AND, $\mathcal{D}_{ln}^{(j)}$}
\label{appendix.B}
Using the orthogonality of the Laguerre polynomials, the following integral identities can be derived:
\begin{eqnarray}
&&\int\!\! \frac{d^2\kper}{(2\pi)^2}e^{-2\alpha_k} L^1_{l-1}(2\alpha_k)L^1_{n-1}(2\alpha_k)\kper^2
= - \frac{(eB)^2}{16\pi}n\delta_{l-1}^{n-1}, \label{eq.A2.Lag1}\\
&&\int\!\! \frac{d^2\kper}{(2\pi)^2}e^{-2\alpha_k} L_{l}(2\alpha_k)L_{n}(2\alpha_k)
= \frac{eB}{8\pi}\delta_{l}^{n}. \label{eq.A2.Lag2}
\end{eqnarray}

Using Eqs.~\eqref{eq.A2.Lag1} and \eqref{eq.A2.Lag1}, we now perform the $d^2\kper$ integrals of Eqs.~\eqref{eq.A1.Bln.j}-\eqref{eq.A1.Dln.j} and obtain
\begin{eqnarray}
\mathcal{B}_{ln}^{(2)} &=& -\frac{(eB)^2}{16\pi}n\delta_{l-1}^{n-1}, \label{eq.A2.B2} \\
\mathcal{C}_{ln}^{(0)} &=& \frac{eB}{8\pi}\FB{\delta_{l}^{n}+\delta_{l-1}^{n-1}}, \label{eq.A2.C0} \\
\mathcal{D}_{ln}^{(0)} &=& -\frac{eB}{8\pi}\FB{\delta_{l}^{n-1} + \delta_{l-1}^{n}}. \label{eq.A2.D0}
\end{eqnarray}
It is to be noted that, the Kronecker delta function having a negative index is always zero (\textit{i.e.} $\delta_{-1}^{-1}=0$). This is due to restrictions on the Laguerre polynomials $L_{-1}(z)=L_{-1}^1(z) =0$ used in defining the magnetized Dirac propagator in Eq.~\eqref{eq.A1.D11}.

%~~~~~~~~~~~~~~~~~~~~~~~~~~~~~~~~~~~~~~~~~~~~~~~~~~~~~~~~~~~~~~~~~~~~~~~~~~~~~~~~~~~~~~~~~~~~~~~~

\bibliographystyle{apsrev4-1}
\bibliography{KuboCond-BMu}

\end{document}